\documentclass[floatfix, twocolumn,showpacs,preprintnumbers,amsmath,amssymb]{revtex4}

\usepackage{graphicx}
\usepackage{dcolumn}
\usepackage{bm}

\begin{document}

\preprint{04.20-q, 04.20 Jb, 98.80-k}

\title{Chaplygin electron gas model}

\author{I. Radinschi}
 \affiliation{%
Department of Physics, "Gh. Asachi" Technical University, Iasi, 700050, Romania
}
 \email{radinschi@yahoo.com}

\author{F. Rahaman}
 \affiliation{Dept. of Mathematics, Jadavpur University, Kolkata-700 032, India}
 \email{farook_rahaman@yahoo.com}

\author{M. Kalam}
 \affiliation{Dept. of Phys., Netaji Nagar College for Women, Regent Estate, Kolkata-700092, India}
 \email{mehedikalam@yahoo.co.in}

\author{K. Chakraborty}
\affiliation{
Dept. of Mathematics, Jadavpur University, Kolkata-700 032, India
}

\date{\today}

\begin{abstract}
We provide a new electromagnetic mass model admitting Chaplygin gas equation
of state. We investigate three specializations, the first characterized by a
vanishing effective pressure, the second provided with a constant effective
density and the third is described by a constant effective pressure $p_{0}$.
For these specializations two particular cases are discussed: (1) $\sigma
e^{\frac{\lambda}{2}}=\sigma_{0}$, and (2) $\sigma e^{\frac{\lambda}{2}}=\sigma_{0}r^{s}$,
where $\sigma$, $\sigma_{0}$, $\lambda$ and $s$ represent
the charge density of the spherical distribution, charge density at the center
of the system, metric potential and a constant, respectively. In addition, for
specialization I, case I we found isotropic coordinate as well as Kretschmann
scalar, and for specialization III, case II two special scenarios have been studied.
\end{abstract}

\maketitle

\section{Introduction}

It is well known in literature that electron, modelled as a spherically
symmetric charged distribution of matter contains negative energy density.
That means electron contains exotic matter. It is still unknown, exotic matter
will follow what type of EOS. So scientific community uses several type of EOS
( namely, Phantom energy, Tracker field, Quintessence , Chaplygin gas EOS etc.
). We describe this exotic nature by Chaplygin gas EOS and looking forward for
a better model of electron. In this way, we want to extend our previous work
\cite{1} concerning a electromagnetic mass model admitting conformal motion, and
study several toy models of electrons.

Chaplygin gas model \cite{2,3,4} is characterized by a negative pressure and
is a perfect fluid with the equation of state $p=-\frac{A}{\rho}$, with $A>0$,
$p$ and $\rho$ the fluid pressure and matter-energy density, respectively in
co-moving reference frame. The cosmological models of the Chaplygin class are
able to describe the transition from a decelerated expansion of the universe
to the present stage, which is characterized by cosmic acceleration. Moreover,
the FRW universe filled with a Chaplygin gas has been studied \cite{5}, and we
notice that some theoretical developments of the model have been performed
\cite{6,7,8,9}. Also, very interesting is the generalized Chaplygin gas
\cite{8} which is characterized by two free parameters $A$ and $\alpha$, with an
equation of state for a barotropic fluid given by $p=-\frac{A}{\rho^{\alpha}}$,
$0<\alpha\leq1$, where the case $\alpha=1$ turns in the original Chaplygin
gas model. Moreover, the generalized Chaplygin gas yields a description of
dark matter and dark energy. An extended case is represented by the
combination of Chaplygin gas model and the dust-like matter \cite{9}. Further, the
Chaplygin gas is the simplest model within the class of tachyon cosmological
models \cite{10}. A lot of work has been done for comparing the Chaplygin model
predictions to the observational data picked up from supernova observations,
cosmic microwave background radiation, etc. In recent time, Chaplygin gas EOS
has become very popular \cite{11,12,13,14,15,16,17,18,19,20,21,22,23,24}, and can be consider the initial point for
fascinating scenarios of the evolution of the universe.

Our aim is to provide several new toy electron gas models of electron, and in
this view we have considered a static spherically symmetric charged perfect
fluid and investigated three specializations: (1) effective pressure
$p_{r}(effective)=8\pi p-E^{2}=0$ and here we discuss two cases, (I) $\sigma
e^{\frac{\lambda}{2}}=\sigma_{0}$ and (II) $\sigma e^{\frac{\lambda}{2}
}=\sigma_{0}r^{s}$; (2) effective density $\rho(effective)=8\pi\rho
+E^{2}=constant=\rho_{0}$ with the cases (I) $\sigma e^{\frac{\lambda}{2}}=\sigma_{0}$
and (II) $\sigma e^{\frac{\lambda}{2}}=\sigma_{0}r^{s}$, and
(III) effective pressure $p_{r}(effective)=8\pi p-E^{2}=constant=p_{0}$ with
the same two particular cases (I) $\sigma e^{\frac{\lambda}{2}}=\sigma_{0}$
and (II) $\sigma e^{\frac{\lambda}{2}}=\sigma_{0}r^{s}$. Here $p$, $\rho$ and
$\sigma$ represent the fluid pressure, usual matter density and charge
density of the spherical distribution. The paper is organized as follows: in
Section 2 the basic equations involved in our calculations are presented, in
Section 3 the details of the calculations and the three specializations of the
model with their particular cases are given. Section 4 is devoted to a summary
of the results and concluding remarks.

\section{Einstein-Maxwell Field Equations and Details of the Model}

The starting point for performing our investigations are the Einstein-Maxwell
field equations and the electromagnetic tensor field for perfect fluid, which
combined with an appropriate specific exotic equation of state (EOS) provide
the basic equations used for elaborating a new Chaplygin electron gas model.

Let us consider the electromagnetic tensor field (EMT) for perfect fluid which
is
\begin{equation}
T_{ik}=(\rho+p)u_{i}u_{k}+pg_{ik},
\end{equation}
with $\rho$ the matter density, $p$ the fluid pressure and $u_{i}$ the
velocity four-vector of a fluid element ($u_{i}u^{i}=1$).

Further, we also have the static spherically symmetric space-time which is
taken as
\begin{equation}
ds^{2}=-e^{\nu(r)}dt^{2}+e^{\lambda(r)}dr^{2}+r^{2}(d\theta^{2}+sin^{2}\theta
d\phi^{2}),
\end{equation}
where the functions of radial coordinate $r$,$\ \nu(r)$ and $\lambda(r)$
represents the metric potentials.

In this context, the Einstein Maxwell field equations are
\begin{equation}
e^{-\lambda}[\frac{\lambda^{\prime}}{r}-\frac{1}{r^{2}}]+\frac{1}{r^{2}}=8\pi\rho+E^{2},
\end{equation}
\begin{equation}
e^{-\lambda}[\frac{1}{r^{2}}+\frac{\nu^{\prime}}{r}]-\frac{1}{r^{2}}=8\pi
p-E^{2},
\end{equation}
\begin{align}
\frac{e^{-\lambda}}{2}\left[\frac{(\nu^{\prime})^{2}}{2}+\nu^{\prime\prime}
-\frac{\lambda^{\prime}\nu^{\prime}}{2}+\frac{({\nu^{\prime}
-\lambda^{\prime}})}{r}\right] =8\pi p+E^{2}.
\end{align}
where $p$, $\rho$ and $E(r)$ represent fluid pressure, matter-energy density
and electric field, respectively.

From eq. (3) we obtain
\begin{equation}
e^{-\lambda}=1-\frac{2M(r)}{r}
\end{equation}
and the electric field can be expressed
\begin{equation}
(r^{2}E)^{\prime}=4\pi r^{2}\sigma e^{\frac{\lambda}{2}}.
\end{equation}
Equation (7) can be also equivalently written in the form
\begin{equation}
E(r)=\frac{1}{r^{2}}\int_{0}^{r}4\pi r^{2}\sigma e^{\frac{\lambda}{2}}
dr=\frac{q(r)}{r^{2}},
\end{equation}
where $q(r)$ represents the total charge of the sphere under consideration.

Here the EOS is taken as a perfect fluid described by the equation of state
\begin{equation}
p=-\frac{A}{\rho},
\end{equation}
where $A>0$, with negative pressure, which corresponds to the Chaplygin gas.

\section{Chaplygin Gas Models and Special Cases}

Now we provide several toy models of electrons.

\textbf{Specialization I}

\textbf{Case - I:}
\begin{equation}
\sigma e^{\frac{\lambda}{2}} = \sigma_{0}
\end{equation}
( $\sigma_{0}$ ia an arbitrary constant ).

Here we assume:
\begin{equation}
p_{r}(effective)=8\pi p-E^{2}=0,
\end{equation}

Now using equations (3)--(9), we get the following solutions as
\begin{equation}
E(r)=\frac{4\pi}{3}\sigma_{0}r,
\end{equation}
\begin{equation}
q(r)=\frac{4\pi}{3}\sigma_{0}r^{3},
\end{equation}
\begin{equation}
p=\frac{2\pi}{9}\sigma_{0}^{2}r^{2},
\end{equation}
\begin{equation}
\rho=-\frac{9A}{2\pi\sigma_{0}^{2}r^{2}},
\end{equation}
\begin{equation}
\nu=2\ln[-81A+4\pi^{2}\sigma_{0}^{4}r^{4}],
\end{equation}
\begin{equation}
e^{-\lambda}=1-\frac{2M(r)}{r},
\end{equation}
where,
\begin{equation}
M(r)=-\frac{18A}{\sigma_{0}^{2}}r+\frac{8\pi^{2}\sigma_{0}^{2}}{45}r^{5}.
\end{equation}

Here one can note that, $p+\rho>0$ (energy condition, limiting case is for
equality with zero) implies
\begin{equation}
a=r>\left[  \frac{81A}{4\pi^{2}\sigma_{0}^{4}}\right]  ^{\frac{1}{4}}
\end{equation}
and

$p+\rho<0$ implies
\begin{equation}
a=r<\left[  \frac{81A}{4\pi^{2}\sigma_{0}^{4}}\right]  ^{\frac{1}{4}}.
\end{equation}

Thus A is restricted due to energy condition. From the expression $\rho
=-\frac{9A}{2\pi\sigma_{0}^{2}r^{2}}$ we observe that the WEC is violated,
only the limiting condition being satisfied for $r\rightarrow\infty$.

We also see that the Kretschmann scalar $R_{\mu\nu\alpha\beta}R^{\mu\nu
\alpha\beta}$ becomes infinity as $r\rightarrow0$.

The assumption of constant charge density can lead to some particular cases.
The central pressure of the fluid sphere $p_{0}$ is zero and $8\pi p=E^{2}$.
From eqs. (12)-(17) it results that a vanishing value of the electric charge
$q$ implies a zero value for $\sigma_{0}$ ($\sigma$), and moreover the
intensity of the electric field and fluid pressure vanish due to a vanishing
charge. Also, this means that we get a new value for the metric potential
$\nu$ and $\sigma_{0}=0$ becomes a singularity point for the matter density
$\rho$, and for the active gravitational mass $M(r)$. We notice that the fluid
pressure is positive with $p\thicksim r^{2}$ and the matter density $\rho$ has
got negative values and $1/r^{2}$ behaviour. The limit $r\rightarrow\infty$
combined with a non-zero value of $\sigma_{0}$ signifies a vanishing value for
the matter density $\rho$, and infinity values for all physical parameters
that appear in eqs. (12)-(14) and (16)-(18). For a radius shrinking to the
center $E(r)$, $q(r)$, $p$ and $M(r) $ vanish, and $\rho$ becomes infinite. In
Fig. 1, Fig.2, Fig.3 and Fig.4 we plot $p$, $\rho$, $e^{-\lambda}$ and $M(r)$
against $r$ parameter.
\begin{figure}
\includegraphics[width=\columnwidth]{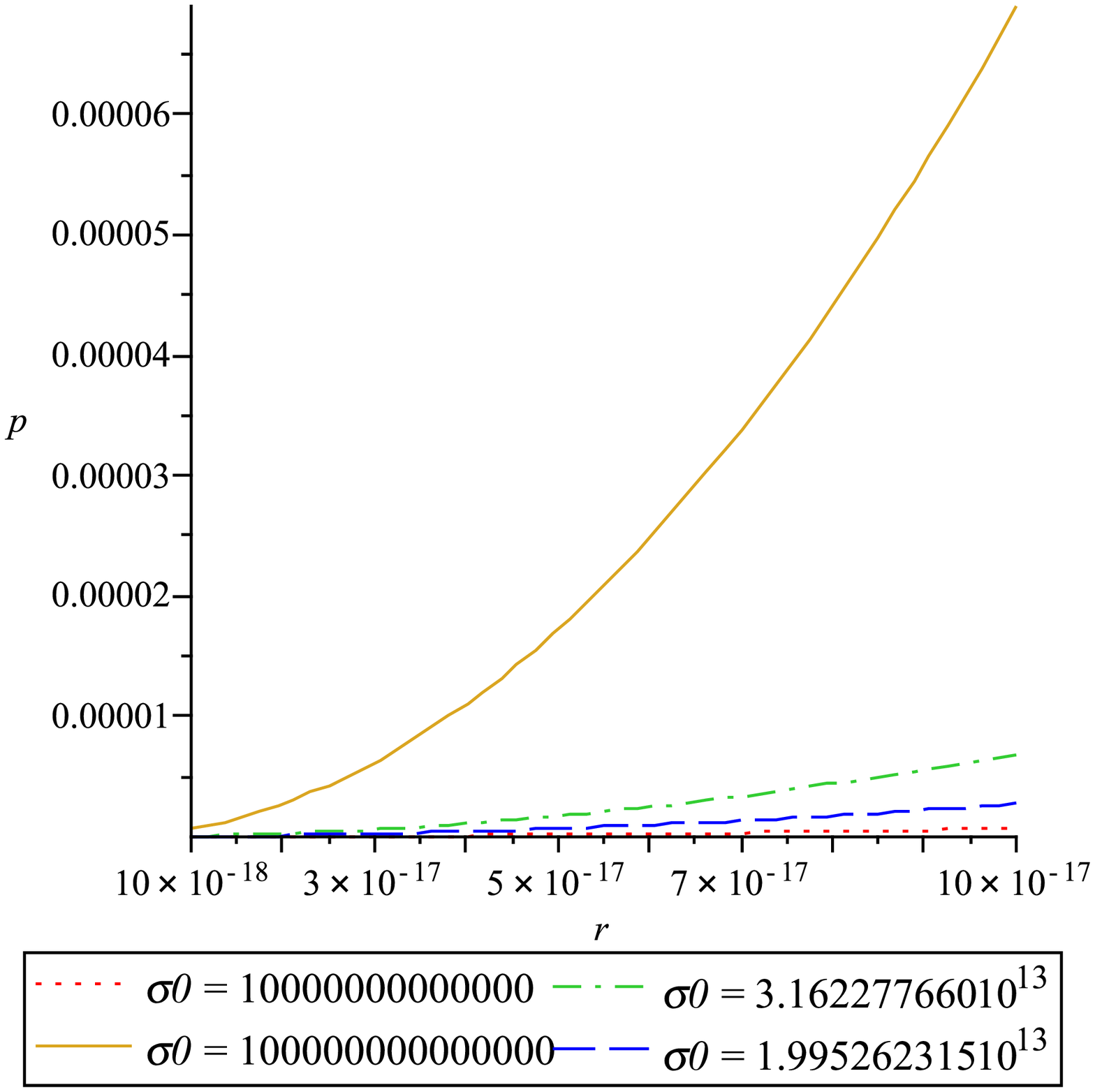}
\caption{}
\label{Fig.1}
\end{figure}
\begin{figure}
\includegraphics[width=\columnwidth]{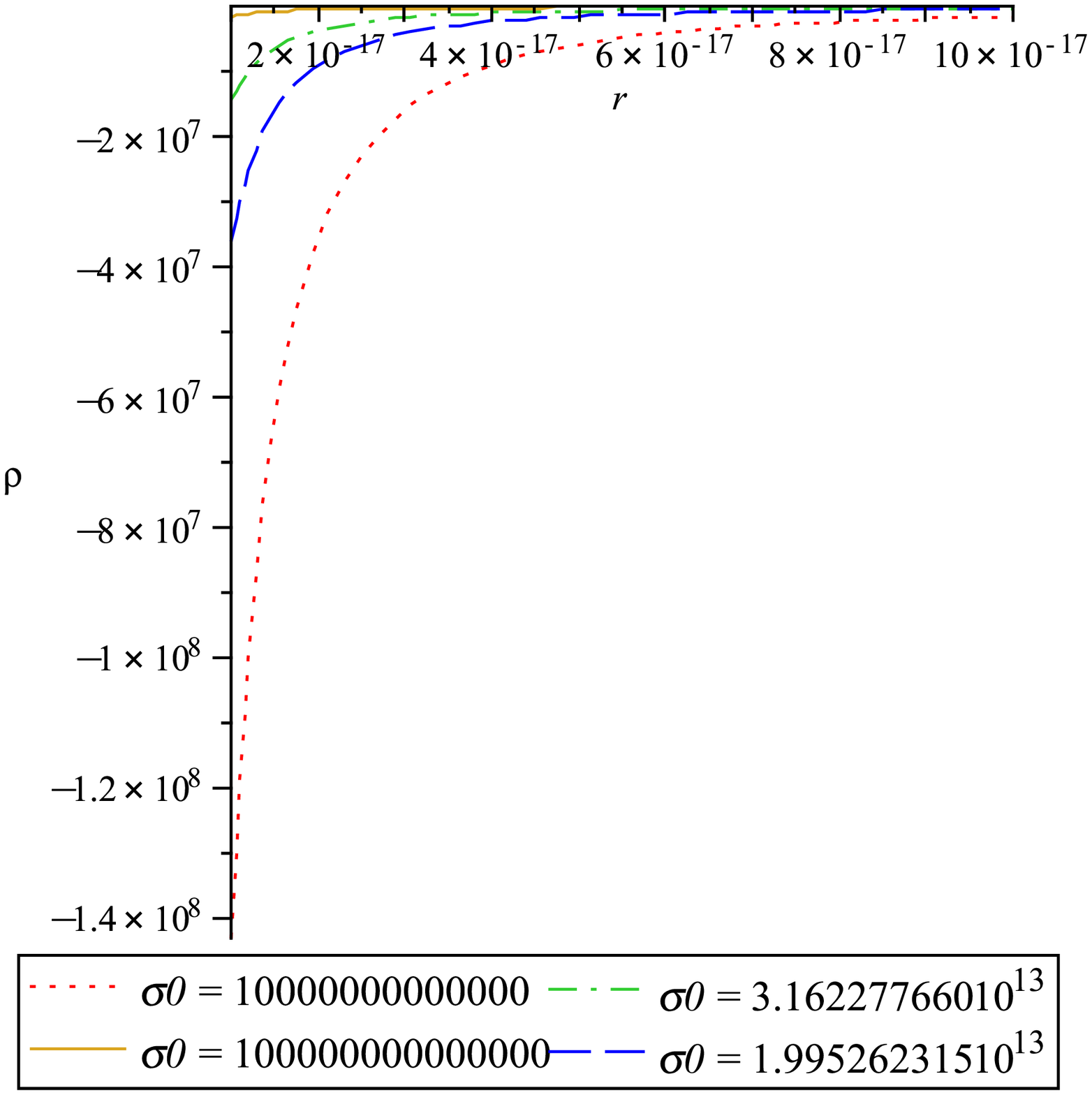}
\caption{}
\label{Fig.2}
\end{figure}
\begin{figure}
\includegraphics[width=\columnwidth]{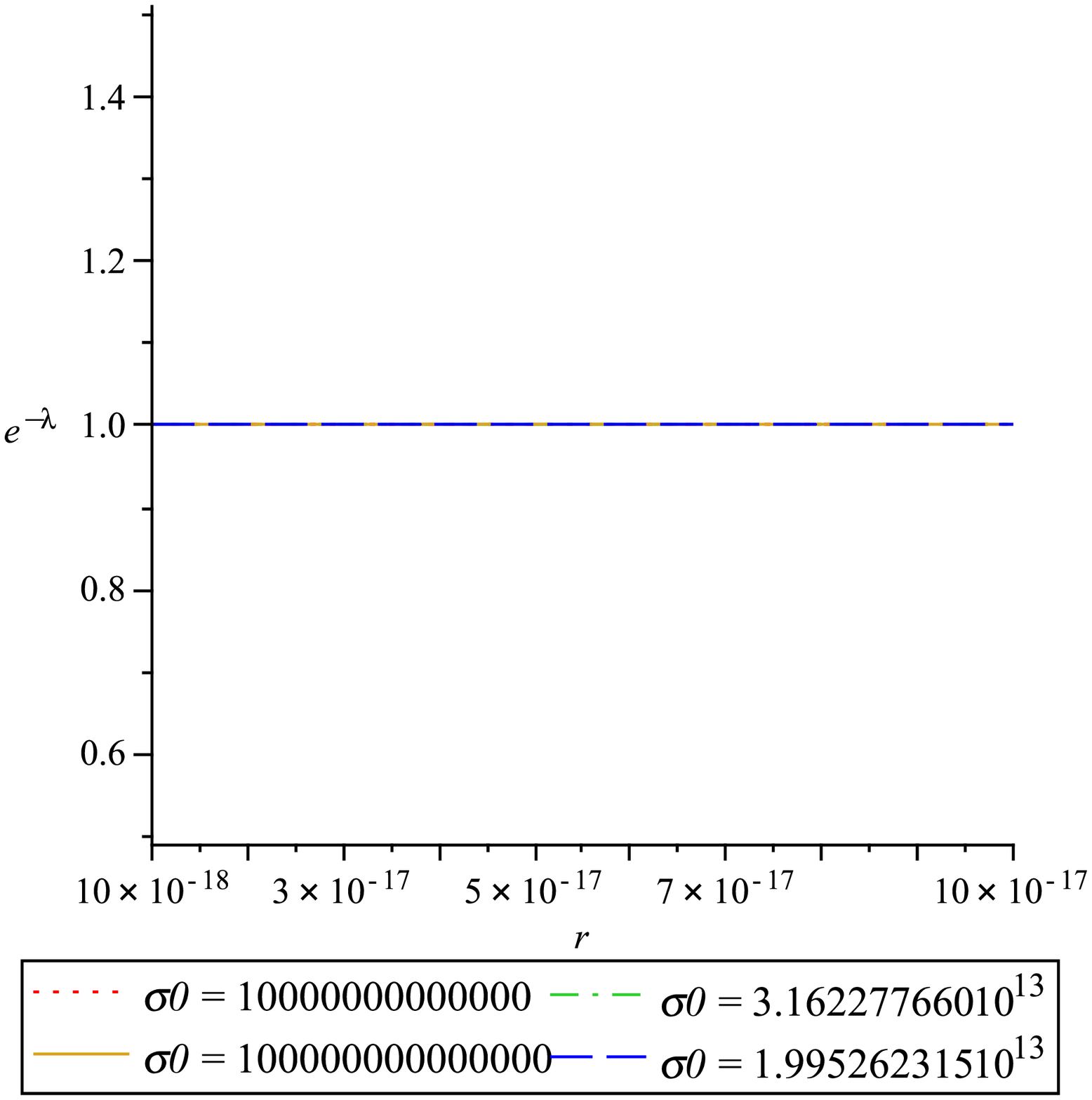}
\caption{}
\label{Fig.3}
\end{figure}
\begin{figure}
\includegraphics[width=\columnwidth]{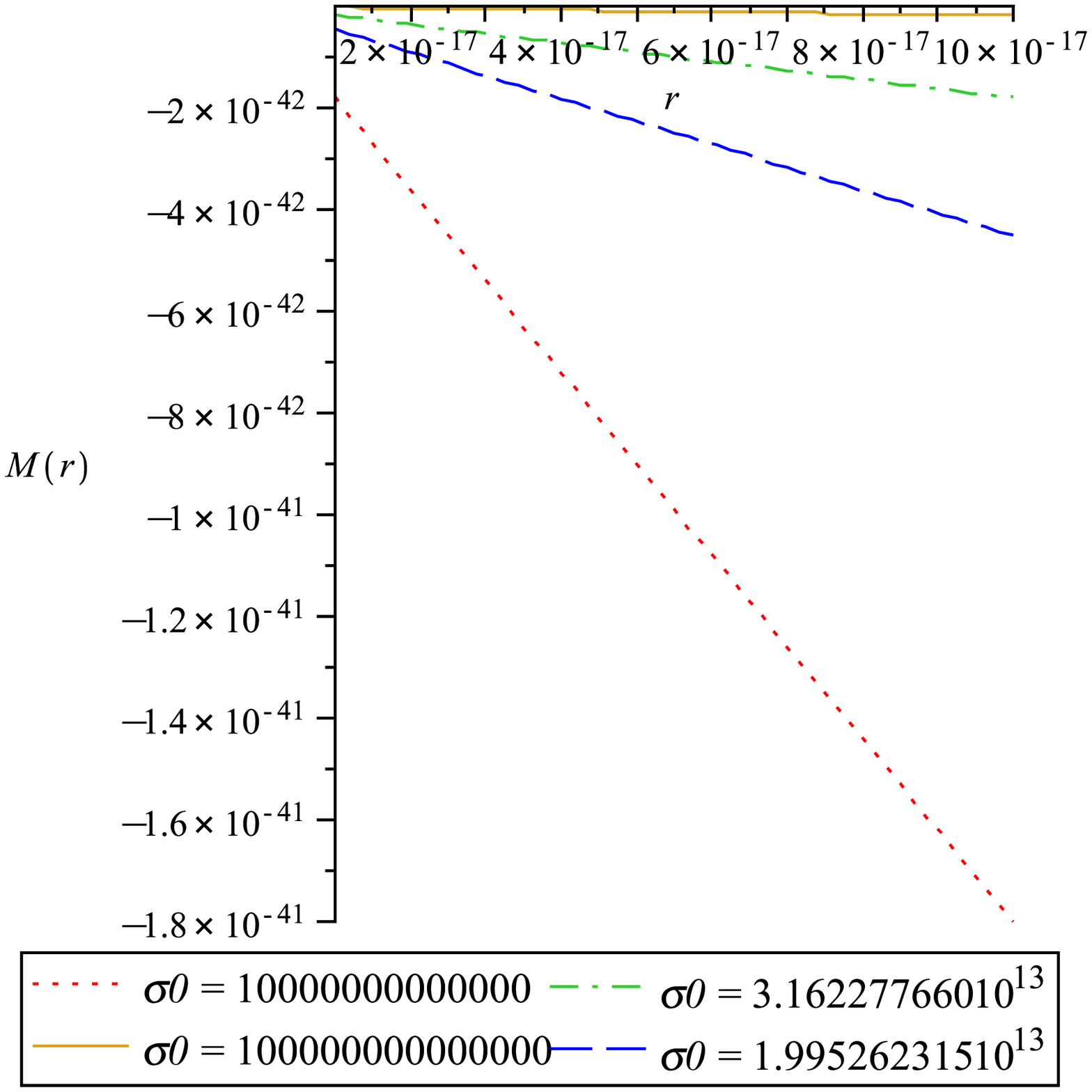}
\caption{\label{Fig.4}}
\end{figure}

To match interior metric with the exterior (Reissner-Nordstr\"{o}m) metric, we
impose only the continuity of $g_{tt}$, $g_{rr}$ and $\frac{\partial g_{tt}
}{\partial r}$, across a surface, S at $r= a $

\begin{equation}
1-\frac{2m}{a}+\frac{Q^{2}}{a^{2}}=[-81A+4\pi^{2}\sigma_{0}^{4}a^{4}]^{2},
\end{equation}
\begin{equation}
1-\frac{2m}{a}+\frac{Q^{2}}{a^{2}}=1+\frac{36A}{\sigma_{0}^{2}}-\frac
{16\pi^{2}\sigma_{0}^{2}a^{4}}{45},
\end{equation}
\begin{equation}
\frac{m}{a^{2}}-\frac{Q^{2}}{a^{3}}=16\pi^{2}\sigma_{0}^{4}a^{3}[-81A+4\pi
^{2}\sigma_{0}^{4}a^{4}].
\end{equation}

From, the above three equations, one could find the values of the unknowns $A
$ and $\sigma_{0}$, $p_{0}$, $\rho_{0}$ ( which of them are occurred in the
cases) in terms of mass, charge and radius of the spherically symmetric
charged objects i.e. electron.

Now, we rewrite our new interior metric in isotropic coordinate as
\begin{equation}
ds^{2}=-e^{\nu}dt^{2}+[\chi(R)]^{2}\left[
\begin{array}
[c]{c}
dR^{2}+\\
+R^{2}(d\theta^{2}+sin^{2}\theta d\phi^{2})
\end{array}
\right].
\end{equation}
Comparing this with our metric (2), one can get
\begin{equation}
r^{2}=\chi^{2}R^{2},
\end{equation}
\begin{equation}
\frac{dr^{2}}{\left(  1+\frac{36A}{\sigma_{0}^{2}}-\frac{16\pi^{2}\sigma
_{0}^{2}}{45}r^{4}\right)  }=\chi^{2}dR^{2}.
\end{equation}

The above two equations yield
\begin{equation}
r^{2}=\frac{2\alpha R^{-\frac{1}{\beta}}}{1+R^{-\frac{2}{\beta}}},
\end{equation}
\begin{equation}
\chi^{2}=\frac{2\alpha R^{-\frac{1}{\beta}-2}}{1+R^{-\frac{2}{\beta}},
}
\end{equation}
[ here, $\alpha^{2}=\frac{45(\sigma_{0}^{2}+36A)}{16\pi^{2}\sigma_{0}^{4}}$
and $\beta=\frac{1}{2\sqrt{1+\frac{36A}{\sigma_{0}^{2}}}}$ ].

Hence, finally, the metric takes the form as $ds^{2}=-\left[  -81A+4\pi
\sigma_{0}^{2}\left(  \frac{2\alpha R^{-\frac{1}{\beta}}}{1+R^{-\frac{2}
{\beta}}}\right)  ^{2}\right]  dt^{2}+\left(  \frac{2\alpha R^{-\frac{1}
{\beta}-2}}{1+R^{-\frac{2}{\beta}}}\right)  \left[  dR^{2}+R^{2}(d\theta
^{2}+sin^{2}\theta d\phi^{2})\right]  $.

In isotropic coordinate system the coordinate singularity at $r=(\frac
{45(1+\frac{36A}{\sigma_{0}^{2}})}{16\pi^{2}\sigma_{0}^{2}})^{\frac{1}{4}}$
has been avoided. We obtain a new metric with metric potentials depending on
$A$, $\sigma_{0}$ and $R$ parameter.

\textbf{Case - II:}

\begin{equation}
\sigma e^{\frac{\lambda}{2}} = \sigma_{0} r^{s}
\end{equation}
($\sigma_{0}$ and s are arbitrary constants )

Here we assume,
\begin{equation}
p_{r} (effective) = 8 \pi p - E^{2} = constant = 0.
\end{equation}

Now using equations (3)--(8) and (17) we get the following solutions as
\begin{equation}
E(r)=\frac{4\pi\sigma_{0}}{(s+3)}r^{s+1},
\end{equation}
\begin{equation}
q(r)=\frac{4\pi\sigma_{0}}{(s+3)}r^{s+3},
\end{equation}
\begin{equation}
p=\frac{2\pi\sigma_{0}^{2}}{(s+3)^{2}}r^{2s+2},
\end{equation}
\begin{equation}
\rho=-\frac{A(s+3)^{2}}{2\pi\sigma_{0}^{2}r^{2+2s}},
\end{equation}
\begin{equation}
\nu=\frac{2}{(s+1)}\ln\left[  -A+\frac{4\pi^{2}\sigma_{0}^{4}}{(s+3)^{4}
}r^{4s+4}\right],
\end{equation}
\begin{equation}
e^{-\lambda}=1-\frac{2M(r)}{r},
\end{equation}
where,
\begin{align}
M(r)  & =\frac{2(s+3)^{2}A}{(2s-1)\sigma_{0}^{2}}r^{-2s+1}+\nonumber \\
& +\frac{8\pi^{2}\sigma_{0}^{2}}{(s+3)^{2}(2s+5)}r^{2s+5}.
\end{align}

Here one can note that, $p+\rho>0$ (energy condition, limiting case is for
equality with zero) implies
\begin{equation}
a=r>\left[  \frac{(s+3)^{4}A}{4\pi^{2}\sigma_{0}^{4}}\right]  ^{\frac{1}
{4s+4}}
\end{equation}
and $p+\rho<0$ implies
\begin{equation}
a=r<\left[  \frac{(s+3)^{4}A}{4\pi^{2}\sigma_{0}^{4}}\right]  ^{\frac{1}
{4s+4}}.
\end{equation}

The expression $\rho=-\frac{A(s+3)^{2}}{2\pi\sigma_{0}^{2}r^{2+2s}}$ yields
only the limiting case of WEC for $s=-3$ or $r\rightarrow\infty$ (for $s>-1$).

Some remarks are needed. A zero value for $\sigma_{0}$ implies vanishing
values for the intensity of the electric field, electric charge and fluid
pressure, a modified value of the metric potential $\nu,$ and is a singularity
for the matter density $\rho$, $e^{-\lambda}$ and active gravitational mass
$M(r)$. For $s=-3$ we get a vanishing value for $\rho$ and and infinity values
for $E(r)$, $q(r)$, $p$, $\nu$, $e^{-\lambda}$ and $M(r)$. \ Concerning the
values $s=-1$, $s=\frac{1}{2}$ and $s=-\frac{5}{2},$ these represent
singularity points for the metric potential $\nu$ and for the gravitational
mass. We observe that the parameters $E(r)$, $q(r)$, $p$, $\rho$, $\nu$ and $M(r)$
have got variations with $r^{s+1}$, $r^{s+3}$, $r^{2\,s+2}$, $1/r^{2\,s+2}$,
ln($r^{4\,s+4})$, $r^{-2\,s+1}$ and $r^{2\,s+5}$, respectively. The value
$s=0$ leads to the case I. At the center of the spherical system $E(r)$,
$q(r)$ and $p$ vanish, the matter density tends to infinity with respect
$s>-1$ and the active gravitational mass becomes infinity for $s>\frac{1}{2}$.

The plots in Fig.5, Fig.6, Fig.7 and Fig.8 display $p$, $\rho$, $e^{-\lambda}$
and $M(r)$ against $r$ parameter for $s=0.4$.
\begin{figure}
\includegraphics[width=\columnwidth]{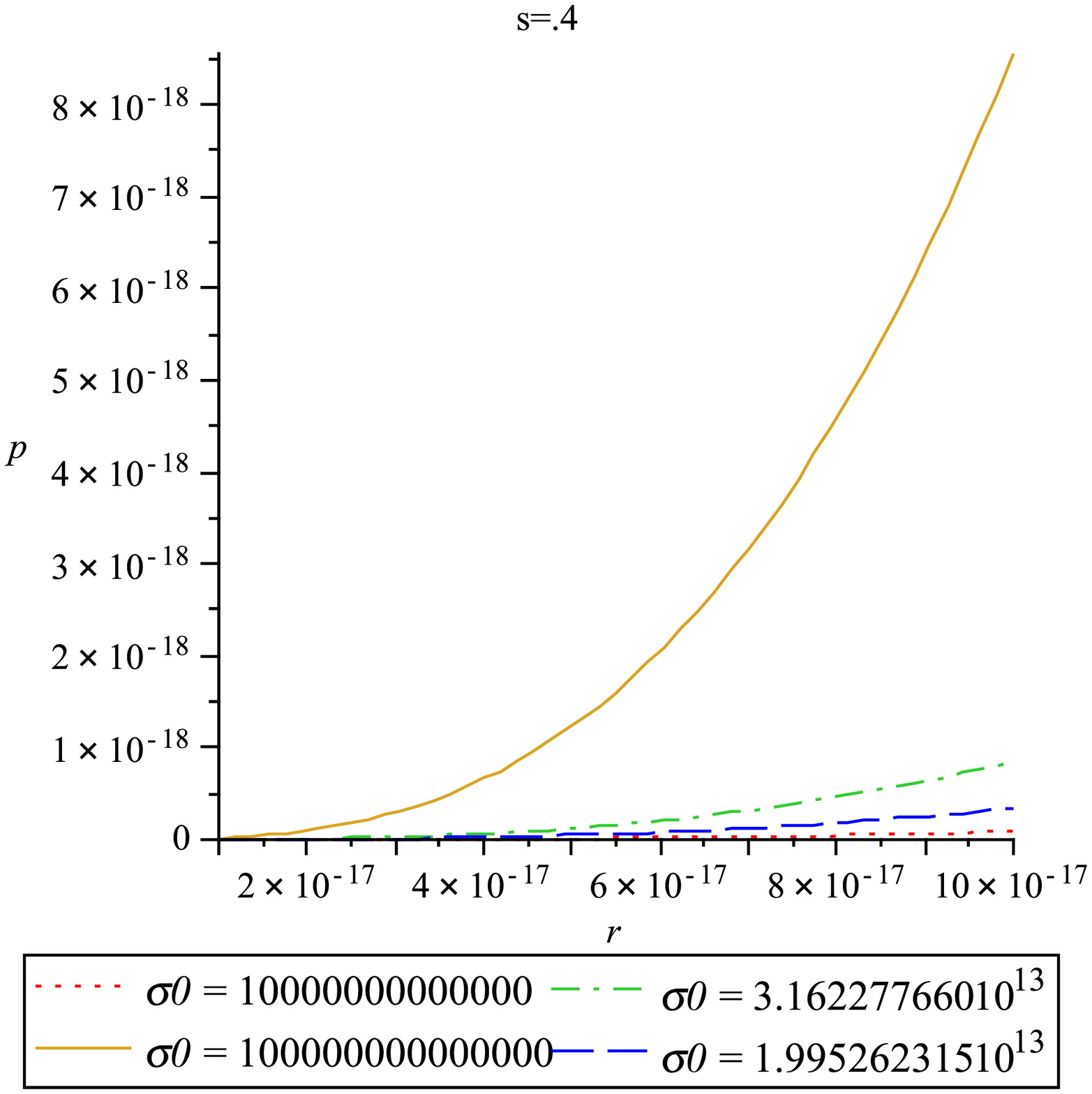}
\caption{\label{Fig.5}}
\end{figure}
\begin{figure}
\includegraphics[width=\columnwidth]{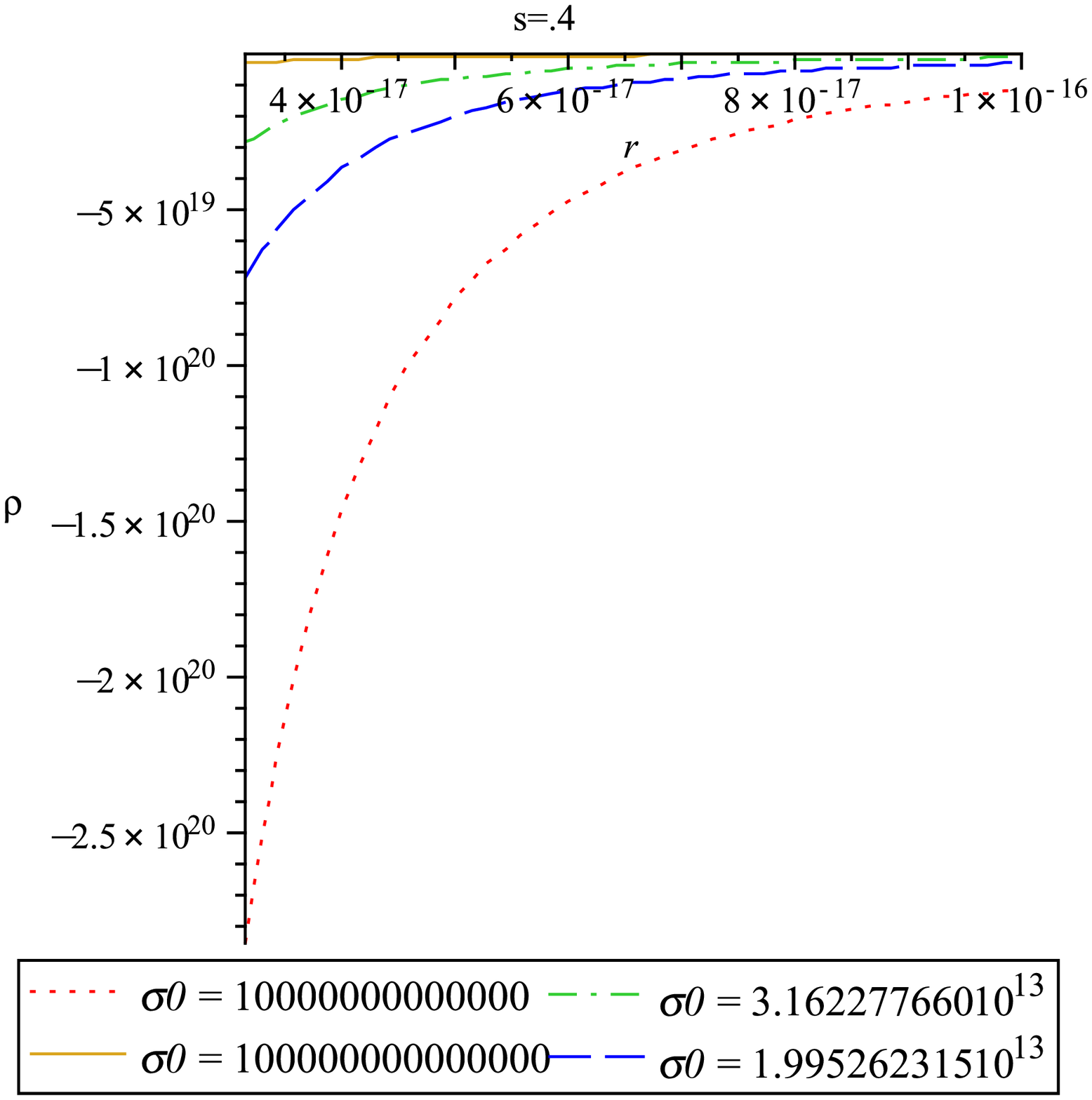}
\caption{\label{Fig.6}}
\end{figure}
\begin{figure}
\includegraphics[width=\columnwidth]{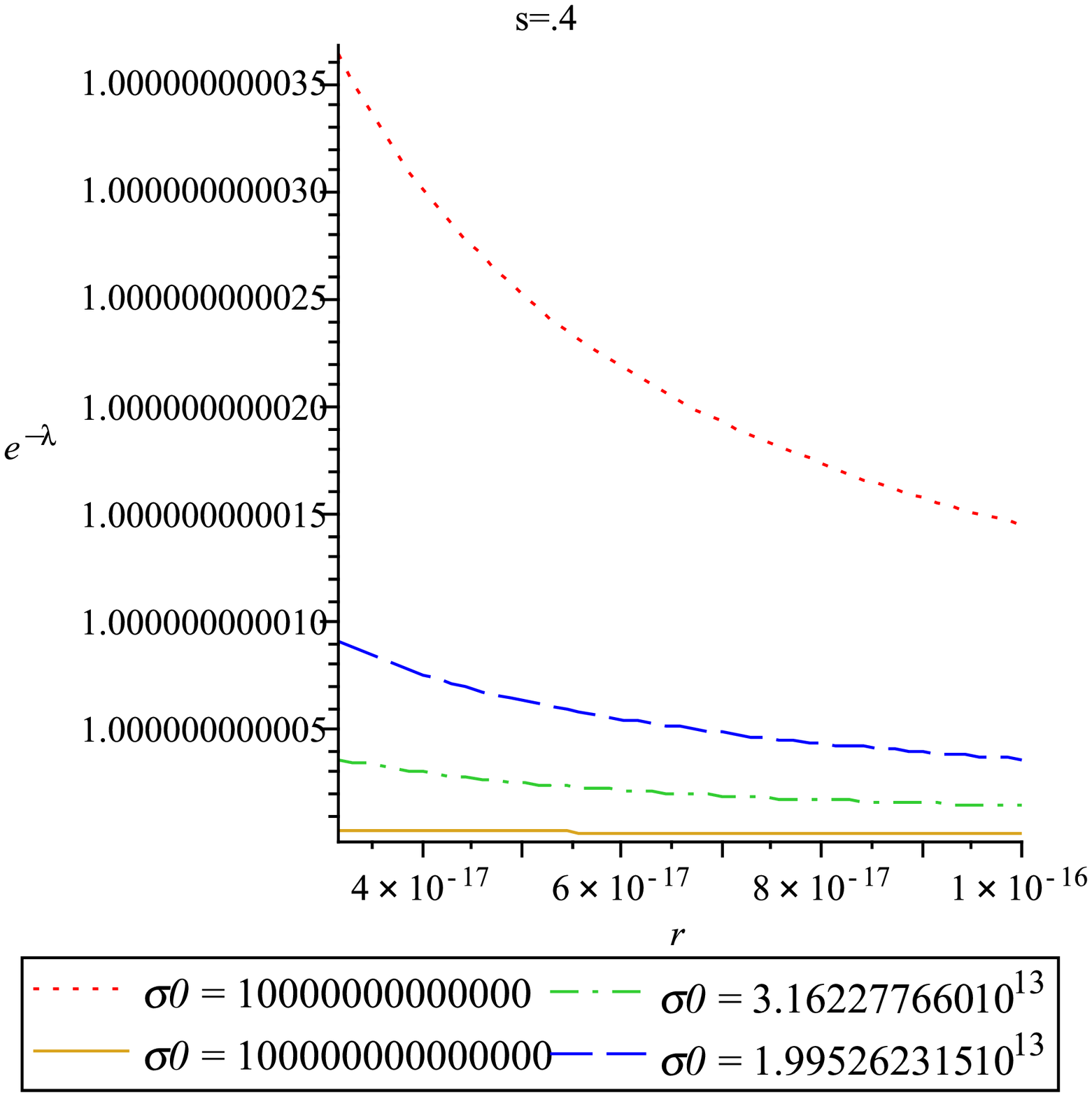}
\caption{\label{Fig.7}}
\end{figure}
\begin{figure}
\includegraphics[width=\columnwidth]{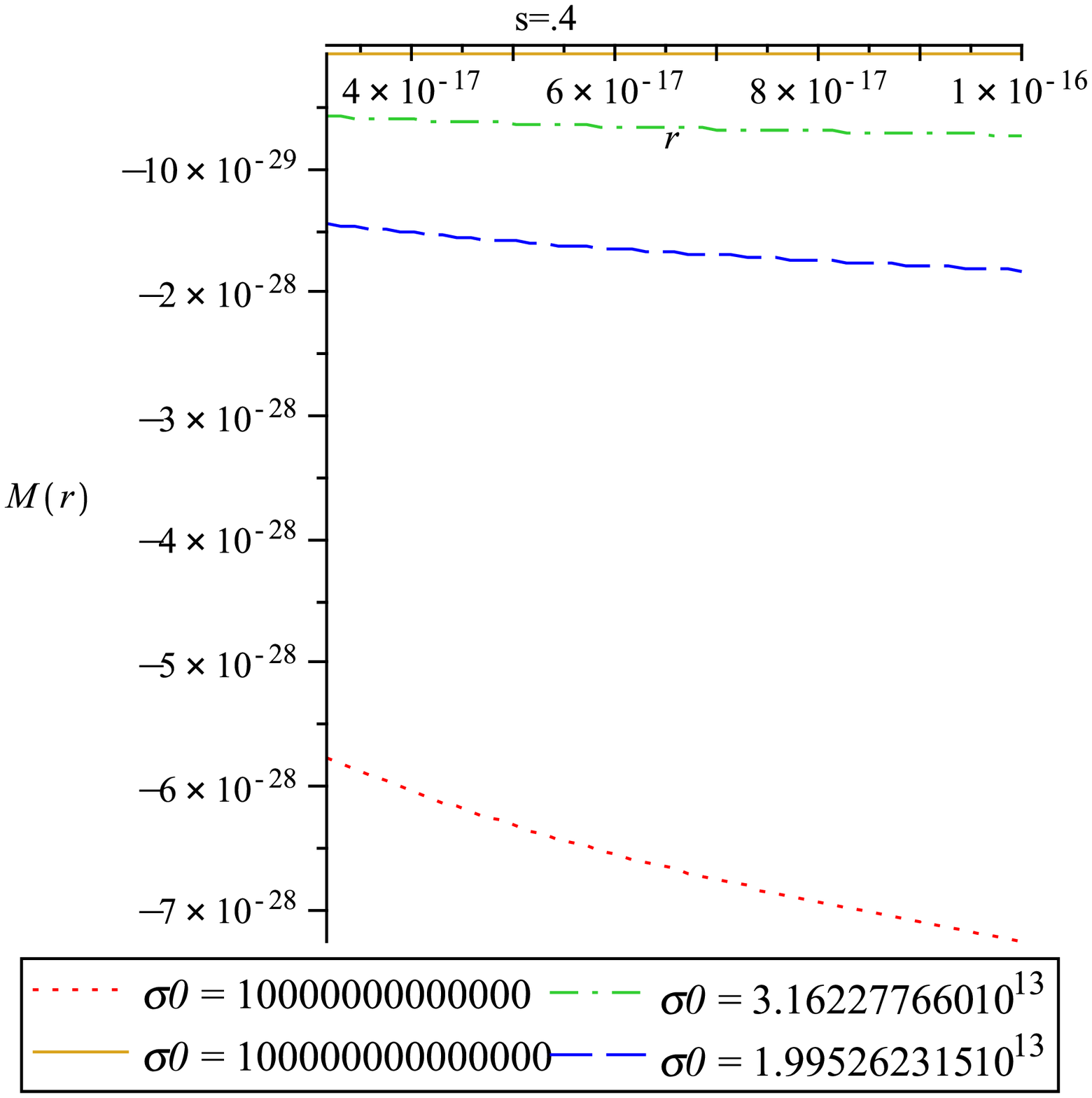}
\caption{\label{Fig.8}}
\end{figure}

Thus $A$ is restricted due to energy condition. To match interior metric with
the exterior (Reissner-Nordstr\"{o}m) metric, we impose only the continuity of
$g_{tt}$, $g_{rr}$ and $\frac{\partial g_{tt}}{\partial r}$, across a surface,
S at $r=a$
\begin{equation}
1-\frac{2m}{a}+\frac{Q^{2}}{a^{2}}=\left[  -A+\frac{4\pi^{2}\sigma_{0}^{4}
}{(s+3)^{4}}r^{4s+4}\right]  ^{\frac{2}{(s+1)}},
\end{equation}
\begin{align}
1-\frac{2m}{a}+\frac{Q^{2}}{a^{2}}  & =1-\frac{4(s+3)^{2}A}{(2s-1)\sigma
_{0}^{2}}a^{-2s}-\nonumber \\
& -\frac{16\pi^{2}\sigma_{0}^{2}}{(s+3)^{2}(2s+5)}a^{2s+4}
\end{align}
\begin{align}
\frac{m}{a^{2}}-\frac{Q^{2}}{a^{3}}  & =\frac{4\pi^{2}\sigma_{0}^{4}
(4s+4)}{(s+1)(s+3)^{4}}a^{4s+3}\cdot \nonumber \\
& \cdot\left[  -A+\frac{4\pi^{2}\sigma_{0}^{4}}{(s+3)^{4}}r^{4s+4}\right]
^{\frac{1-s}{(s+1)}}.
\end{align}
These equations yield the values of $A$ and $\sigma_{0}$, $p_{0}$, $\rho_{0}$
( which of them are occurred in the cases) in terms of mass, charge and radius
of the spherically symmetric charged objects i.e. electron.

\textbf{Specialization II}

Here we assume,
\begin{equation}
\rho(effective)=8\pi\rho+E^{2}=constant=\rho_{0}.
\end{equation}

\textbf{Case - I:}
\begin{equation}
\sigma e^{\frac{\lambda}{2}} = \sigma_{0}
\end{equation}
( $\sigma_{0}$ ia an arbitrary constant )

Here, the solutions are
\begin{equation}
\rho= \frac{1}{8\pi}\left[  \rho_{0} - \frac{16 \pi^{2}}{9} \sigma_{0}^{2}
r^{2} \right],
\end{equation}
\begin{equation}
p = \frac{-8\pi A}{\left[  \rho_{0} - \frac{16 \pi^{2}}{9} \sigma_{0}^{2}
r^{2} \right] },
\end{equation}
\begin{equation}
\nu=2\ln\left[  \frac{\left[  \rho_{0}-\frac{16\pi^{2}}{9}\sigma_{0}^{2}
r^{2}\right]  }{\left[  64A\pi^{2}-\left(  \rho_{0}-\frac{16\pi^{2}}{9}
\sigma_{0}^{2}r^{2}\right)  ^{2}\right]  ^{2}}\right],
\end{equation}
\begin{equation}
e^{-\lambda}=1-\frac{2M(r)}{r},
\end{equation}
where,
\begin{equation}
M(r)=\frac{\rho_{0}}{6}r^{3}.
\end{equation}

We notice that for $8\pi\rho=-E^{2}$ the constant effective density term
$\rho_{0}$ vanishes. This implies that the matter density $\rho$ and fluid
pressure $p$ have new expressions $\rho=-\frac{2\pi}{9}\sigma_{0}^{2}r^{2} $
and $p=\frac{9A}{2\pi\sigma_{0}^{2}r^{2}}$. The matter density $\rho$ is
negative and the fluid pressure $p$ takes positive values according to the
parameters $A$, $\sigma_{0}$ and $r$, respectively. Further, the metric
potential $\nu$ has a new expression and the gravitational mass $M(r)$
vanishes. This implies $e^{-\lambda}=1$ and determines a changing in the
metric given by (2) where the $g_{rr}$ component becomes constant. For
$8\pi\rho\neq-E^{2}$ the matter density $\rho$ is positive and the fluid
pressure $p$ is negative with the condition $\rho_{0}>\frac{16\pi^{2}}
{9}\sigma_{0}^{2}r^{2}$. The negativity of the gravitational mass $\rho
_{0}<-\frac{16\pi^{2}}{3}\sigma_{0}^{2}a^{2}$ and energy density ($\rho<0$,
$\rho_{0}<\frac{16\pi^{2}}{9}\sigma_{0}^{2}a^{2}$) in the case of the electron
is connected with the Reissner-Nordstr\"{o}m repulsion. The case
$r\rightarrow0$ leads to $\rho=\frac{1}{8\pi}\rho_{0}$ which is constant,
$\nu=2\ln\left[  \frac{\rho_{0}}{\left[  64A\pi^{2}-\rho_{0}^{2}\right]  ^{2}
}\right]  $ and the fluid pressure takes negative values according to
$p=-\frac{8\pi A}{\rho_{0}}$, the same expressions as those obtained for
$\sigma_{0}=0$. Further, the quantity $e^{-\lambda}=1$ and $M(r)$ is equal to zero.

In Fig.9, Fig.10, Fig.11 and Fig.12 we have the graphs for $8\pi\rho$, $8\pi
p$, $e^{-\lambda}$ and $M(r)$ against $r$, respectively.
\begin{figure}
\includegraphics[width=\columnwidth]{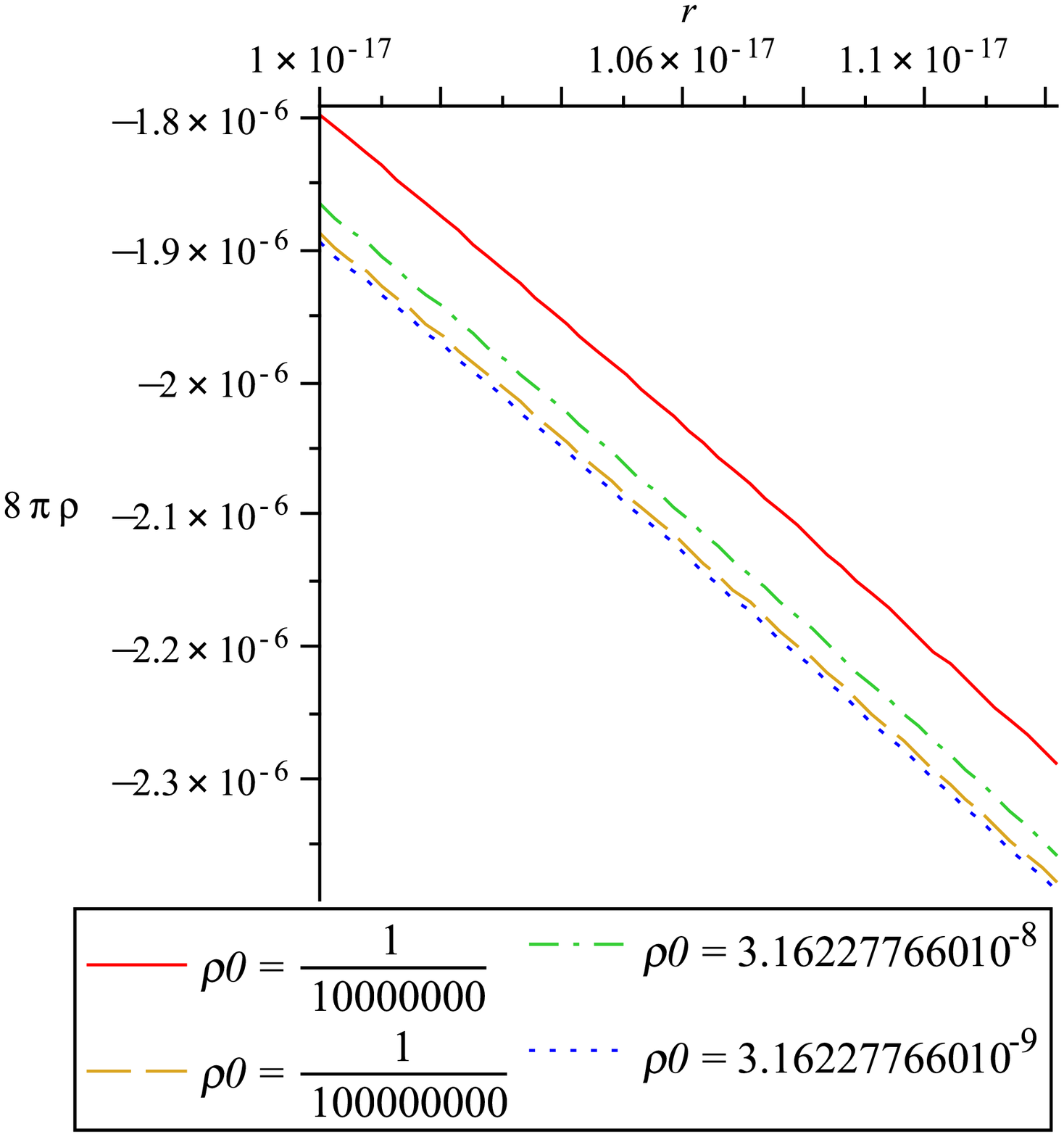}
\caption{\label{Fig.9}}
\end{figure}
\begin{figure}
\includegraphics[width=\columnwidth]{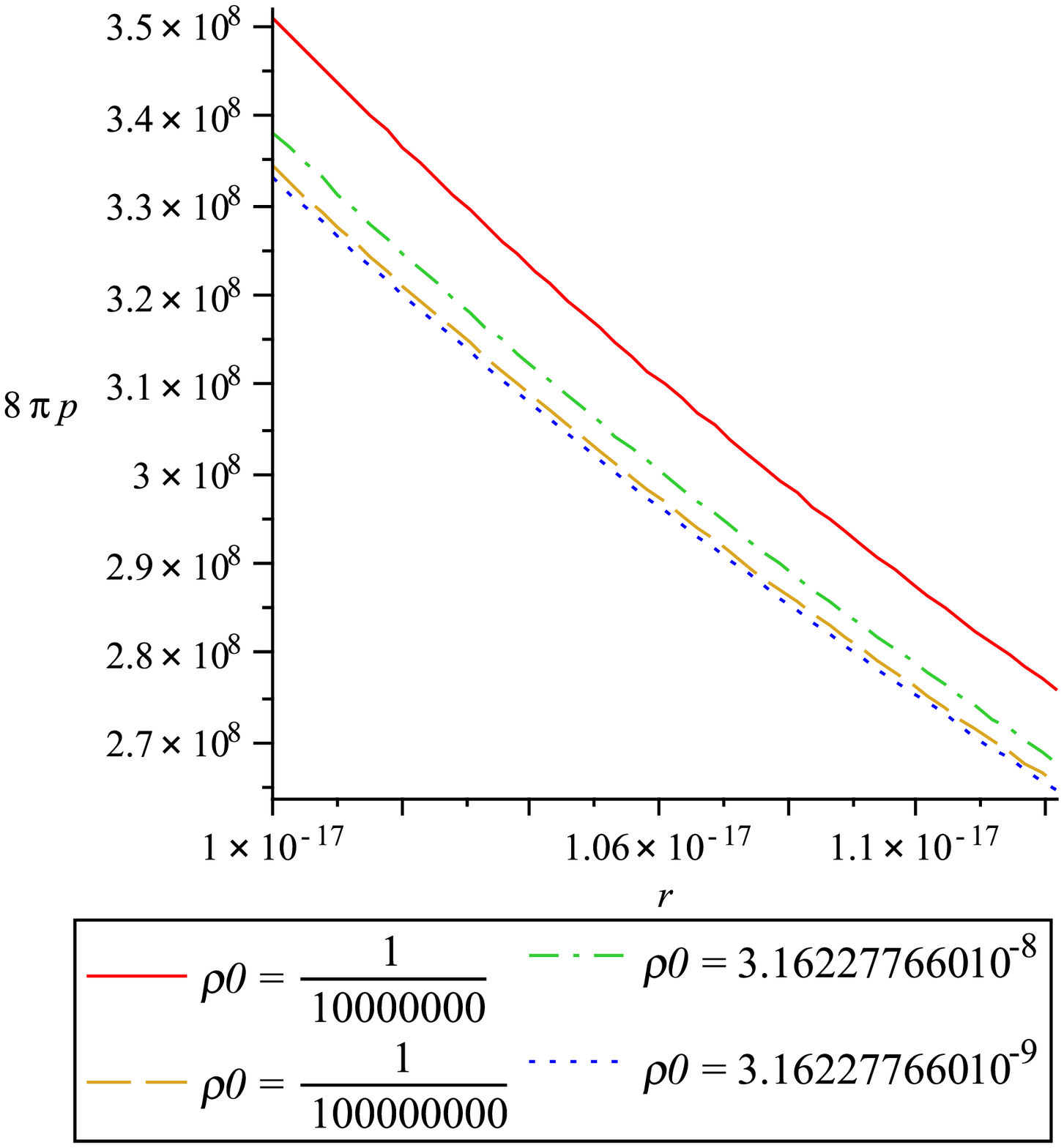}
\caption{\label{Fig.10}}
\end{figure}
\begin{figure}
\includegraphics[width=\columnwidth]{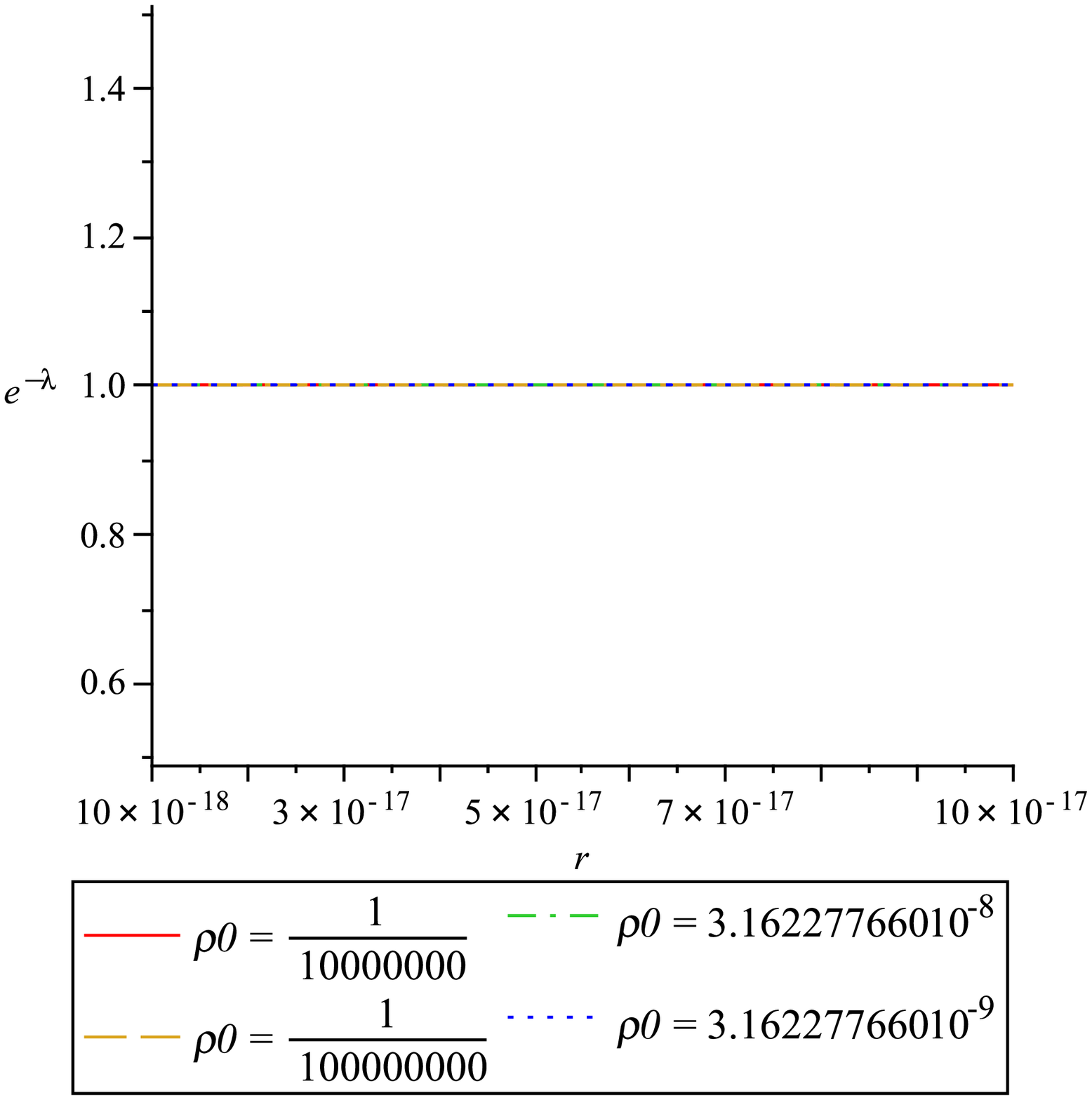}
\caption{\label{Fig.11}}
\end{figure}
\begin{figure}
\includegraphics[width=\columnwidth]{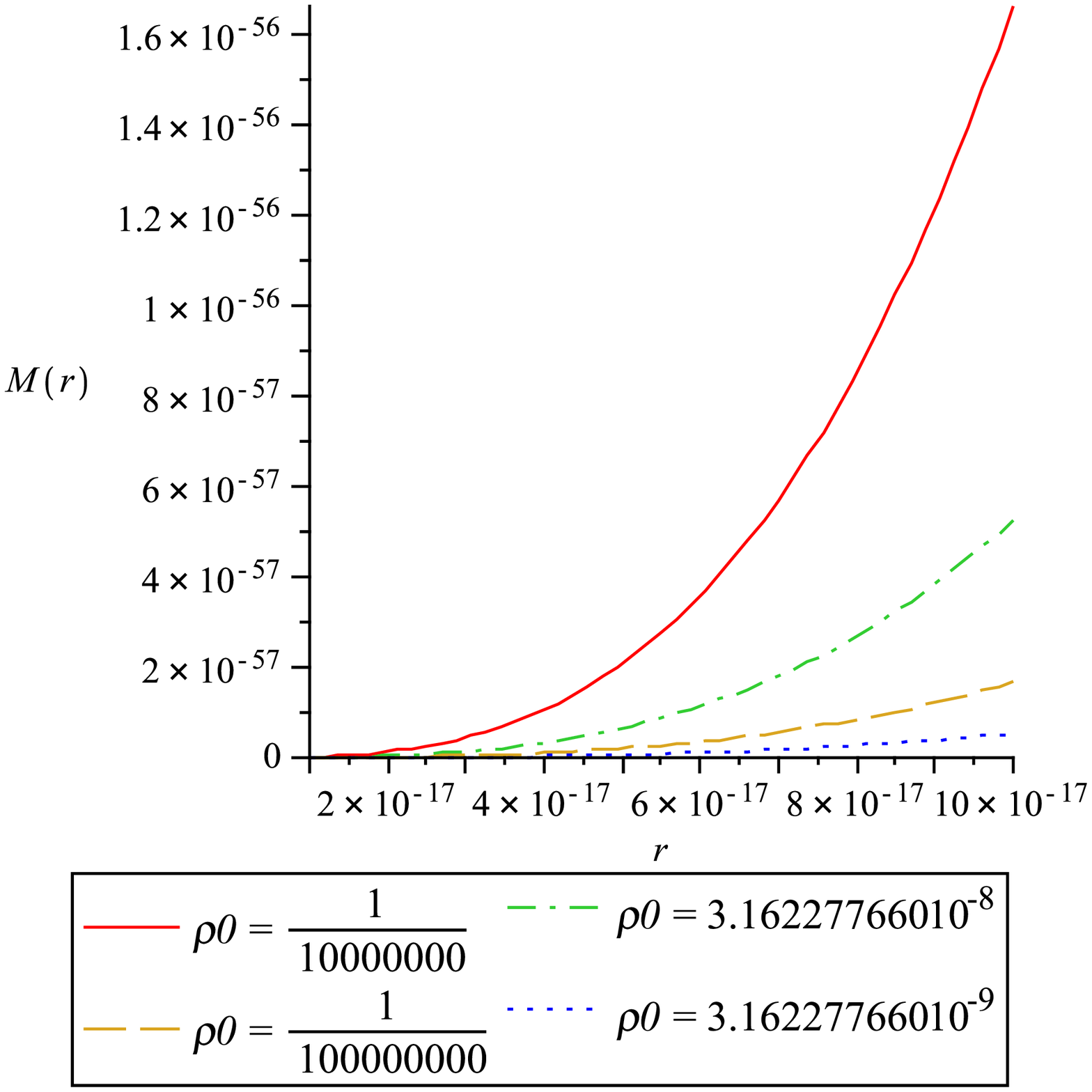}
\caption{\label{Fig.12}}
\end{figure}

From $\rho=\frac{1}{8\pi}\left[  \rho_{0}-\frac{16\pi^{2}}{9}\sigma_{0}
^{2}r^{2}\right]  $ it results that $\rho_{0}\geqslant\frac{16\pi^{2}}
{9}\sigma_{0}^{2}r^{2}$ corresponds to WEC. Also, $\rho_{0}-\frac{16\pi^{2}
}{9}\sigma_{0}^{2}r^{2}>\sqrt{64A\pi^{2}}$ (energy condition, limiting case is
for equality with zero) or $<\sqrt{64A\pi^{2}}$ .

\textbf{Case - II:}
\begin{equation}
\sigma e^{\frac{\lambda}{2}} = \sigma_{0} r^{s}
\end{equation}
($\sigma_{0}$ and $s$ are arbitrary constants )
Here we assume,
\begin{equation}
\rho(effective)=8\pi\rho+E^{2}=constant=\rho_{0},
\end{equation}

Here, the solutions are
\begin{equation}
\rho= \frac{1}{8\pi}\left[  \rho_{0} - \frac{16 \pi^{2}\sigma_{0}^{2}
}{(s+3)^{2}} r^{2s+2} \right],
\end{equation}
\begin{equation}
p = \frac{-8\pi A}{\left[  \rho_{0} - \frac{16 \pi^{2}\sigma_{0}^{2}
}{(s+3)^{2}} r^{2s+2}\right] },
\end{equation}
\begin{equation}
\nu=2\ln\left[  \frac{\left[  \rho_{0}-\frac{16\pi^{2}\sigma_{0}^{2}
}{(s+3)^{2}}r^{2s+2}\right]  }{\left[  64A\pi^{2}-\left(  \rho_{0}-\frac
{16\pi^{2}\sigma_{0}^{2}}{(s+3)^{2}}r^{2s+2}\right)  ^{2}\right]  ^{2}
}\right],
\end{equation}
\begin{equation}
e^{-\lambda}=1-\frac{2M(r)}{r},
\end{equation}
where,
\begin{equation}
M(r)=\frac{\rho_{0}}{6}r^{3}.
\end{equation}

Here also for $8\pi\rho=-E^{2}$ the constant effective density term $\rho_{0}$
vanishes and $\rho$, $p$, $\nu$ and $e^{-\lambda}$ have finite values and the
active gravitational mass is zero. With $8\pi\rho\neq-E^{2} $, as in the case
I, for $r=0$ all the physical parameters given by eqs. (52)-(54) are nonzero
finite quantities (the same as for $\sigma_{0}=0$). For $s=0$ we have the
physical meaning of Case I. Also, $e^{-\lambda}=1$ and $M(r)$ vanishes. For a
non-vanishing $\rho_{0}$ the fluid pressure is negative for $\rho_{0}
>\frac{16\pi^{2}\sigma_{0}^{2}}{(s+3)^{2}}r^{2s+2}$.

In Fig.13, Fig.14, Fig.15 and Fig.16 we plot $8\pi\rho$, $8\pi p$,
$e^{-\lambda}$ and $M(r)$ against $r$ for $s=0.05$.
\begin{figure}
\includegraphics[width=\columnwidth]{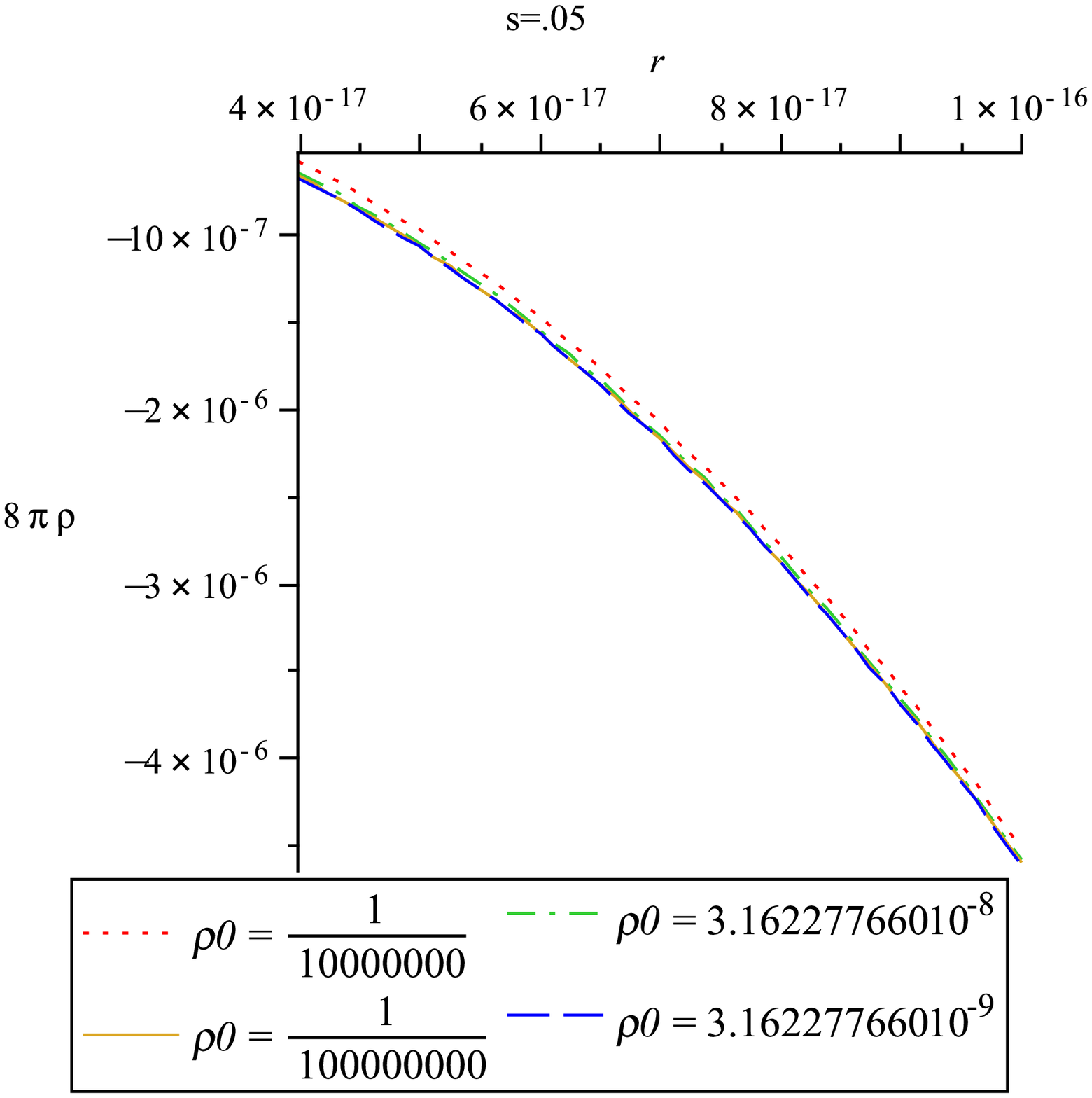}
\caption{\label{Fig.13}}
\end{figure}
\begin{figure}
\includegraphics[width=\columnwidth]{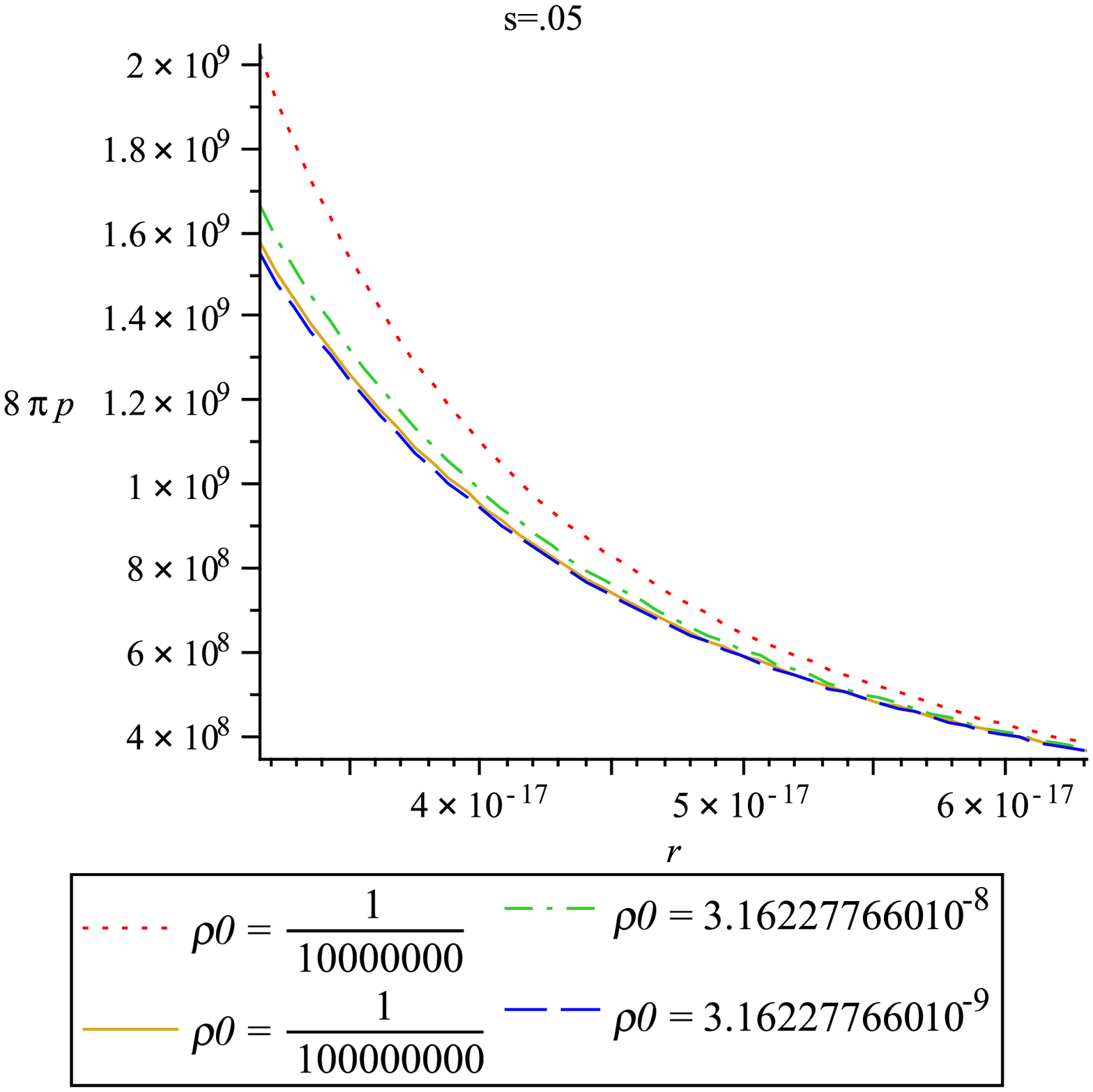}
\caption{\label{Fig.14}}
\end{figure}
\begin{figure}
\includegraphics[width=\columnwidth]{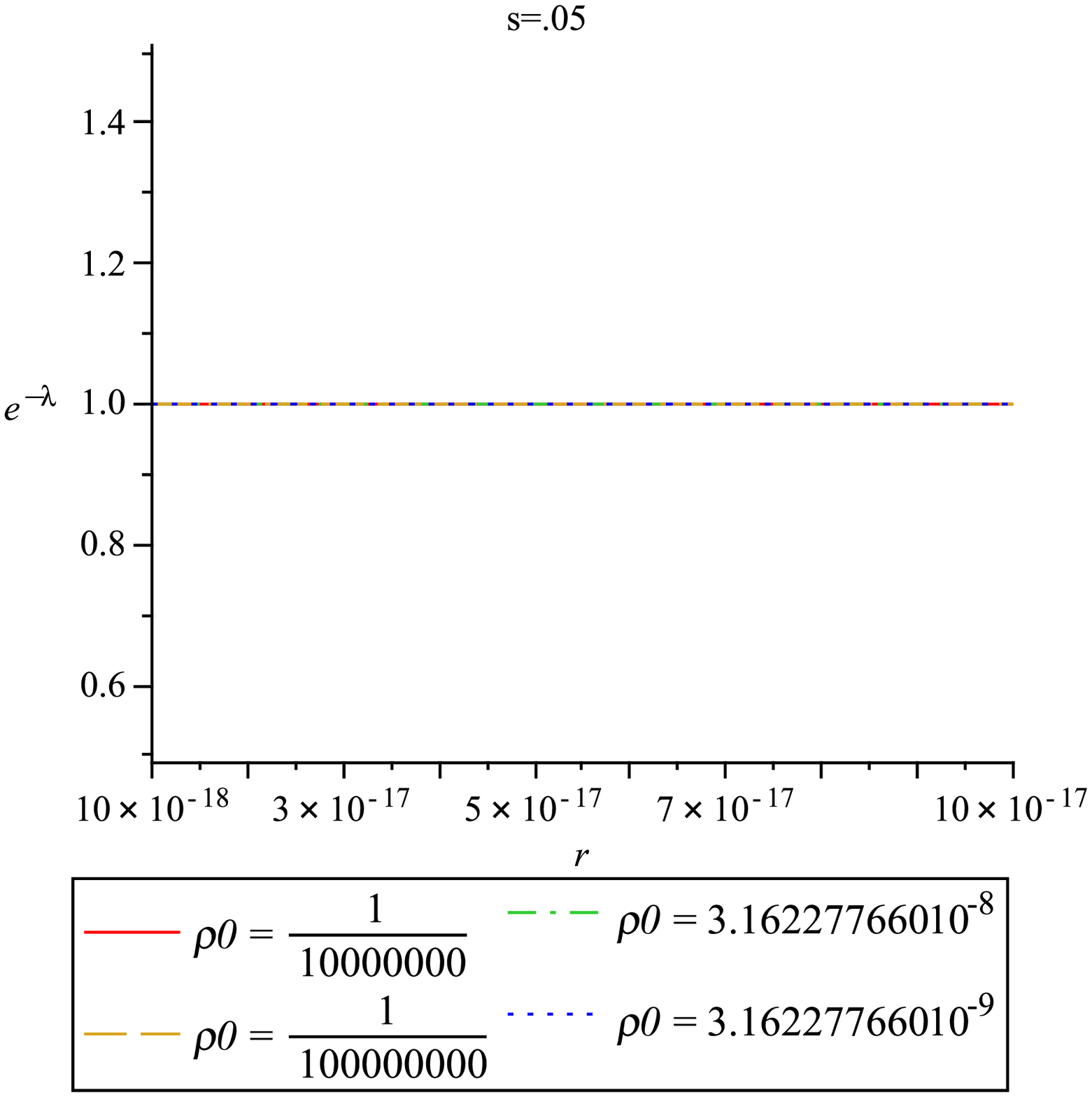}
\caption{\label{Fig.15}}
\end{figure}
\begin{figure}
\includegraphics[width=\columnwidth]{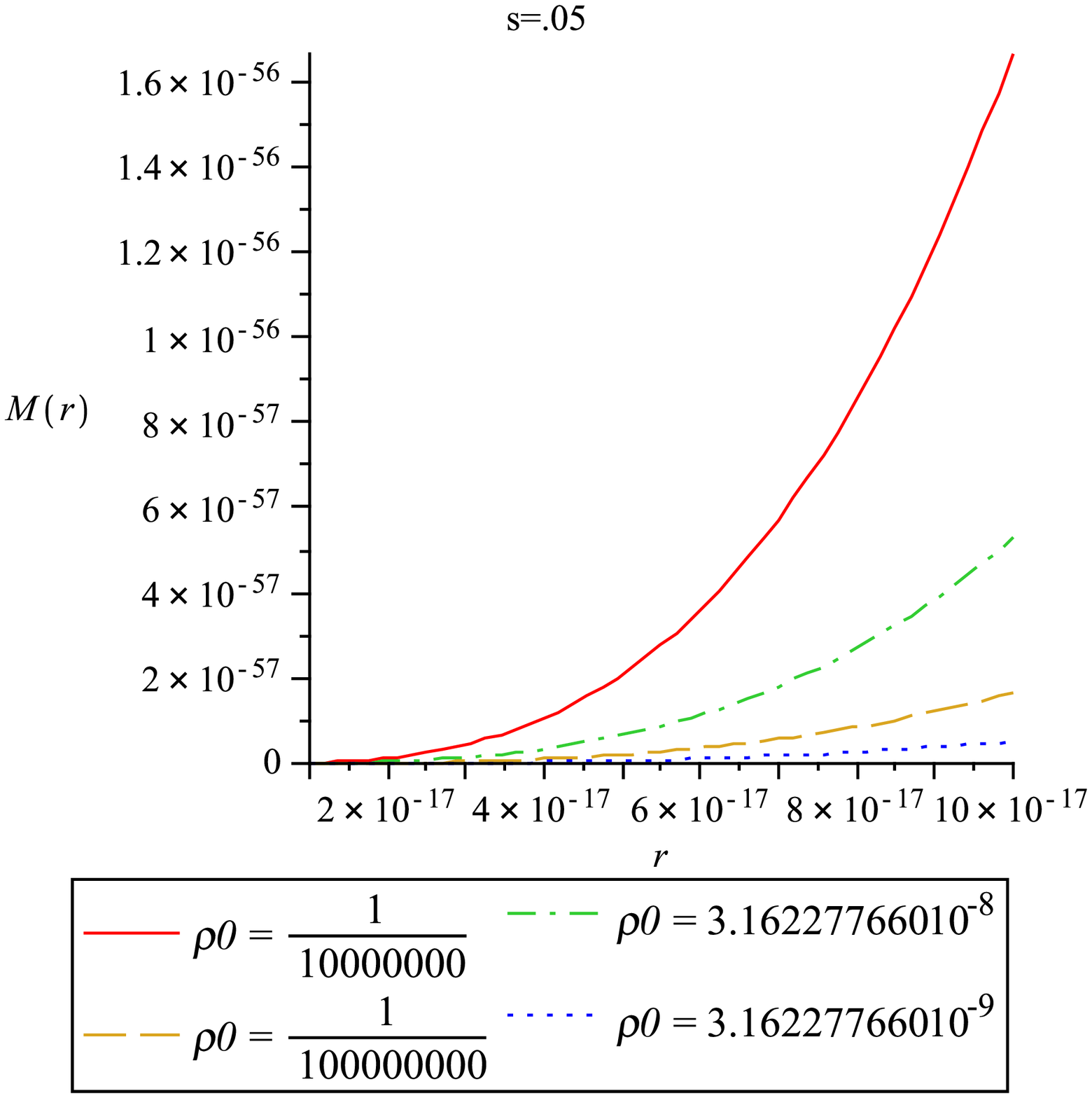}
\caption{\label{Fig.16}}
\end{figure}

From $\rho=\frac{1}{8\pi}\left[  \rho_{0}-\frac{16\pi^{2}\sigma_{0}^{2}
}{(s+3)^{2}}r^{2s+2}\right]  $ we note that for $\rho_{0}\geqslant\frac
{16\pi^{2}\sigma_{0}^{2}}{(s+3)^{2}}r^{2s+2}$ WEC is satisfied. We have
$\rho_{0}-\frac{16\pi^{2}\sigma_{0}^{2}}{(s+3)^{2}}r^{2s+2}>\sqrt{64A\pi^{2}}$
(energy condition, limiting case is for equality with zero) or $<\sqrt
{64A\pi^{2}}$.

\newpage
\textbf{Specialization III}

\textbf{Case - I:}
\begin{equation}
\sigma e^{\frac{\lambda}{2}} = \sigma_{0}
\end{equation}
( $\sigma_{0}$ ia an arbitrary constant )
Here we assume,
\begin{equation}
p_{r}(effective)=8\pi p-E^{2}=constant=p_{0},
\end{equation}
Here, the solutions are
\begin{equation}
p = \frac{1}{8\pi}\left[  p_{0} + \frac{16 \pi^{2}}{9} \sigma_{0}^{2} r^{2}
\right],
\end{equation}
\begin{equation}
\rho= \frac{-8\pi A}{\left[  p_{0} + \frac{16 \pi^{2}}{9} \sigma_{0}^{2} r^{2}
\right] },
\end{equation}
\begin{equation}
\nu=2\ln\left[  -5184A\pi^{2}+(9p_{0}+16\pi^{2}\sigma_{0}^{2}r^{2}
)^{2}\right],
\end{equation}
\begin{equation}
e^{-\lambda}=1-\frac{2M(r)}{r},
\end{equation}
where,
\begin{widetext}
\begin{eqnarray}
M(r)=4\pi\left[  \frac{2\pi\sigma_{0}^{2}}{45}r^{5}-
\frac{9A}{2\pi\sigma_{0}^{2}}\left(
r-\sqrt{\frac{9p_{0}}{16\pi^{2}\sigma_{0}^{2}}}tan^{-1}\sqrt{\frac{16\pi
^{2}\sigma_{0}^{2}}{9p_{0}}}r
\right)  \right].
\end{eqnarray}
\end{widetext}
To match interior metric with the exterior (Reissner-Nordstr\"{o}m) metric, we
impose only the continuity of $g_{tt}$, $g_{rr}$ and $\frac{\partial g_{tt}
}{\partial r}$, across a surface, S at $r=a$
\begin{widetext}
\begin{equation}
1-\frac{2m}{a}+\frac{Q^{2}}{a^{2}}=\left[
-5184A\pi^{2}+(9p_{0}+16\pi^{2}\sigma_{0}^{2}a^{2})^{2}
\right]  ^{2},
\end{equation}
\begin{eqnarray}
1-\frac{2m}{a}+\frac{Q^{2}}{a^{2}}= 1-\frac{8\pi}{a}\times
\left[
\frac{2\pi\sigma_{0}^{2}}{45}a^{5}
-\frac{9A}{2\pi\sigma_{0}^{2}}\left(  a-\sqrt{\frac{9p_{0}}{16\pi^{2}
\sigma_{0}^{2}}}tan^{-1}\sqrt{\frac{16\pi^{2}\sigma_{0}^{2}}{9p_{0}}}a\right)
\right],
\end{eqnarray}
\begin{align}
\frac{m}{a^{2}}-\frac{Q^{2}}{a^{3}}  & =64(9p_{0}+16\pi^{2}\sigma_{0}^{2}
a^{2})\pi^{2}\sigma_{0}^{2}a\cdot \left[  -5184A\pi^{2}+(9p_{0}+16\pi^{2}\sigma_{0}^{2}a^{2})^{2}\right].
\end{align}
\end{widetext}
With the above three equations, one could determine the values of the unknowns
$A$ and $\sigma_{0}$, $p_{0}$, $\rho_{0}$ expressed in terms of mass, charge
and radius of the spherically symmetric charged objects i.e. electron.

At $r=0$, $p$, $\rho$, $e^{\nu}$ and $e^{\lambda}$ are all nonzero finite
quantities i.e. there is no singularity at $r=0$\textbf{. }We point out that
for $8\pi p=E^{2}$ the constant effective pressure $p_{0}$ vanishes. We found
for the fluid pressure, matter density and $M(r)$\ the same expressions as for
specialization I, case I. The metric potential $\nu$\ becomes $\nu=2\ln\left[
-5184A\pi^{2}+(16\pi^{2}\sigma_{0}^{2}r^{2})^{2}\right]  $. In the case $8\pi
p\neq E^{2}$ there is an upper limit for the value of $p_{0}$ imposed by the
condition of positivity of the gravitational mass. For $p_{0}<-\frac{16\pi
^{2}}{9}\sigma_{0}^{2}r^{2}$ the fluid pressure takes negative values and the
matter density is positive (WEC, $\rho\neq0$ no limiting WEC). The fluid
pressure vanishes for $p_{0}=-\frac{16\pi^{2}}{9}\sigma_{0}^{2}r^{2}$.

Here, $p+\rho$ $> $ or $< 0$, implies

$p_{0}+\frac{16\pi^{2}\sigma_{0}^{2}}{9}a^{2}$ $>\sqrt{64A\pi^{2}}$ (energy
condition, limiting case is for equality with zero) or $<\sqrt{64A\pi^{2}}$.
From the expression $\rho=\frac{-8\pi A}{\left[  p_{0}
+\frac{16\pi^{2}}{9}\sigma_{0}^{2}r^{2}\right]  }$ it results the violation of
the limiting case of WEC, condition which satisfied only for $r\rightarrow
\infty$.

The plots from Fig.17, Fig.18, Fig.19 and Fig.20 display $8\pi p$, $8\pi\rho$,
$e^{-\lambda}$ and $M(r)$ against $r$.
\begin{figure}[!h]
\includegraphics[width=\columnwidth]{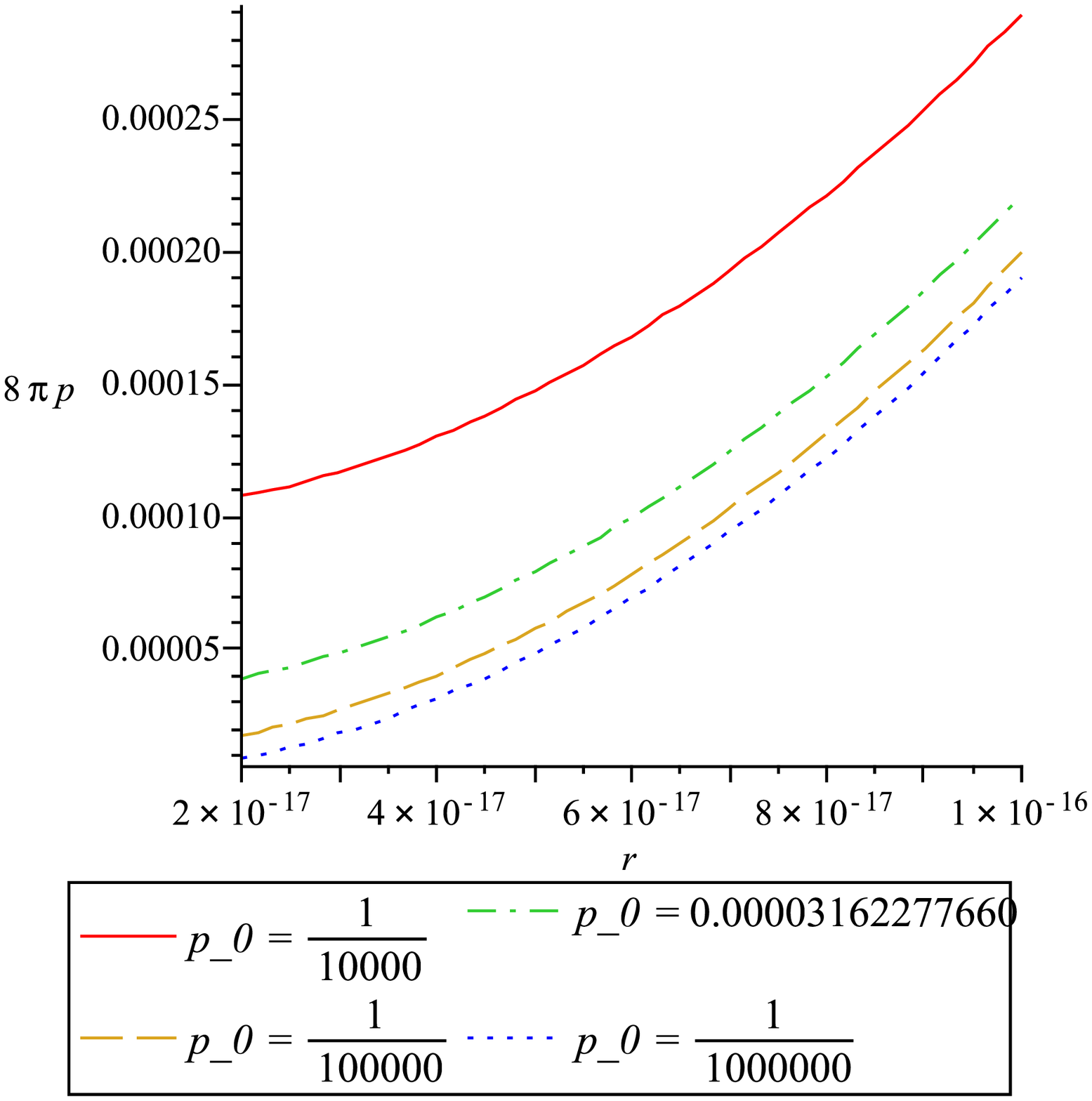}
\caption{\label{Fig.17}}
\end{figure}
\begin{figure}[!h]
\includegraphics[width=\columnwidth]{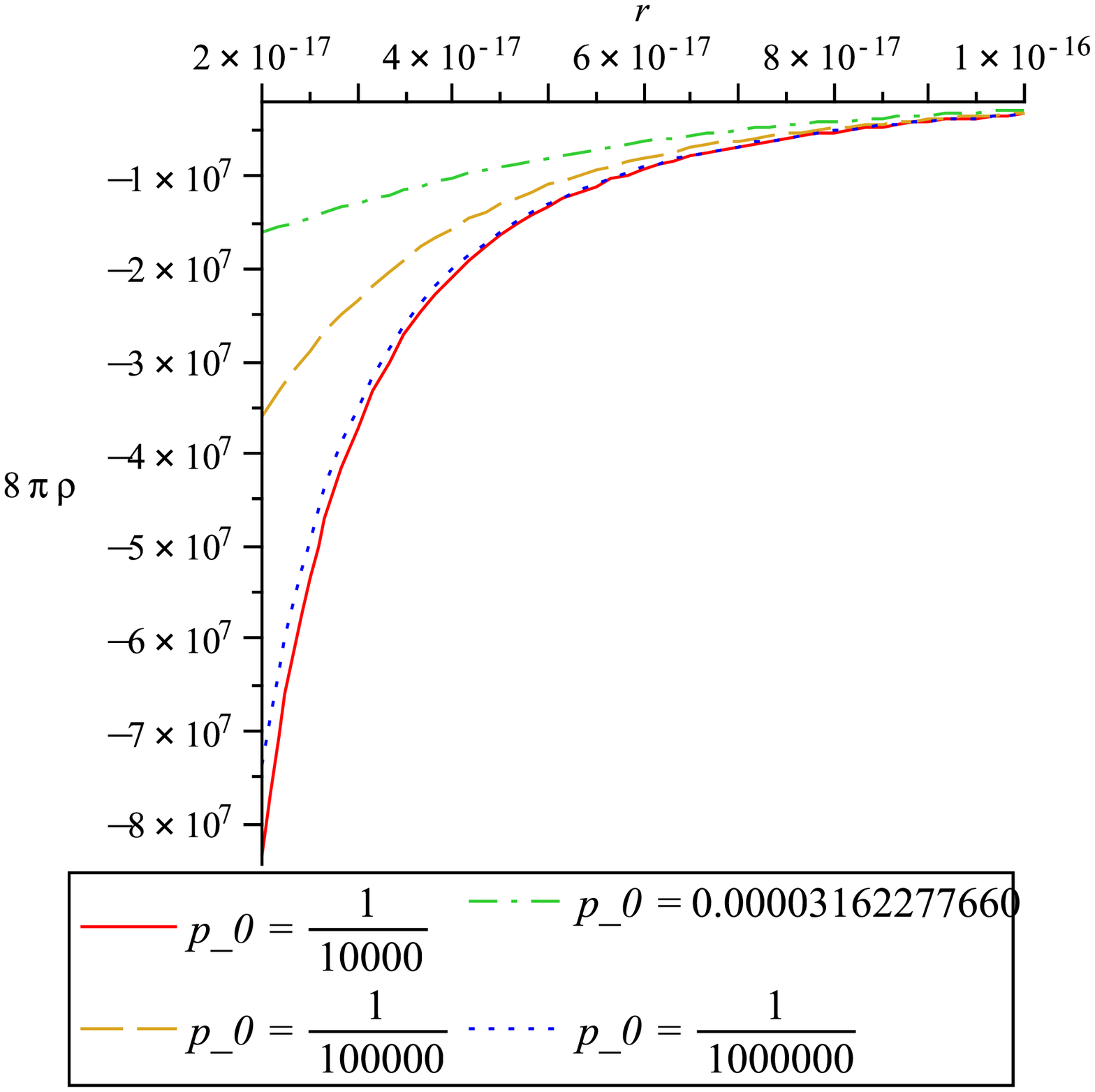}
\caption{\label{Fig.18}}
\end{figure}
\begin{figure}[!t]
\includegraphics[width=\columnwidth]{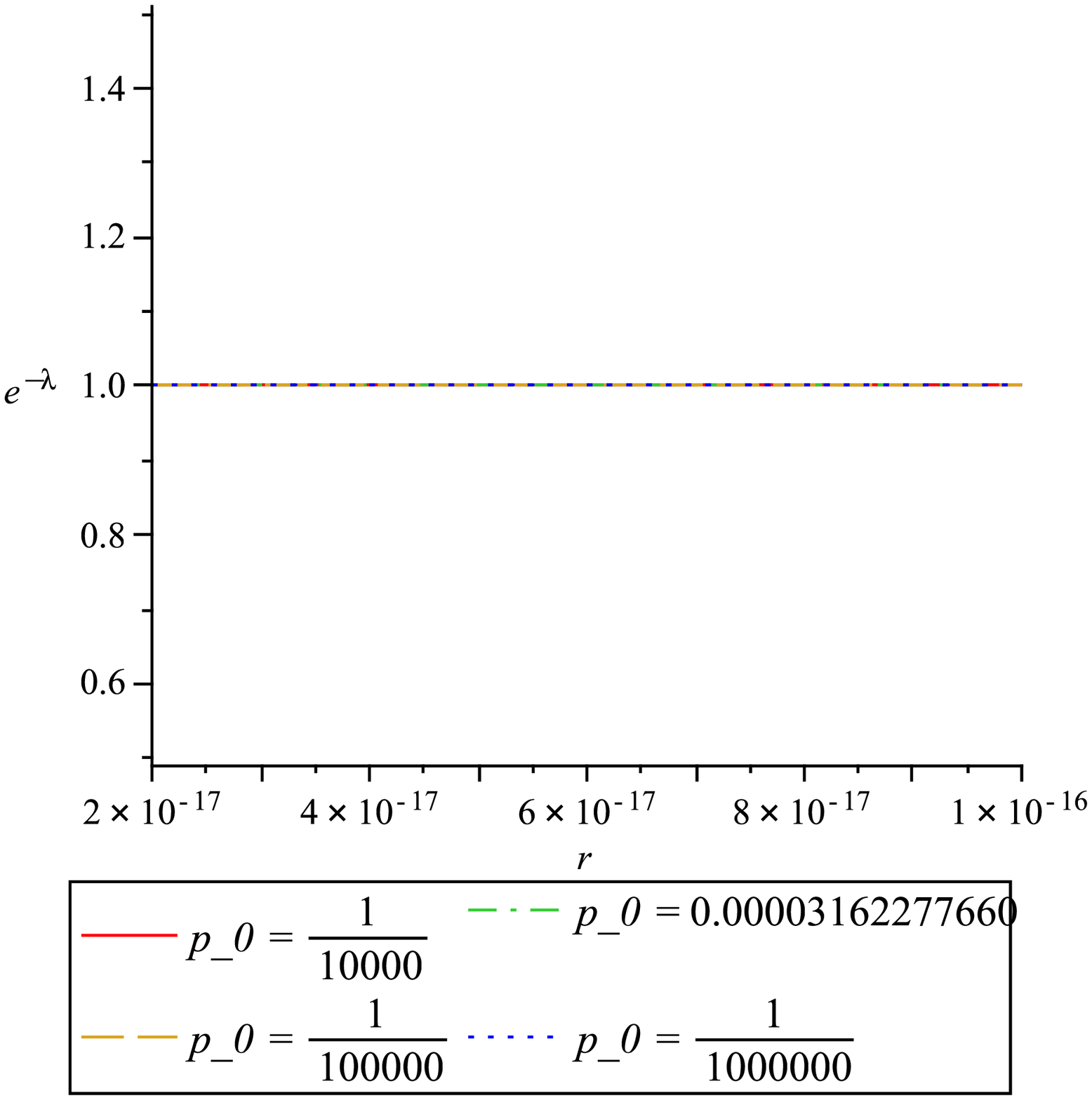}
\caption{\label{Fig.19}}
\end{figure}
\begin{figure}[!t]
\includegraphics[width=\columnwidth]{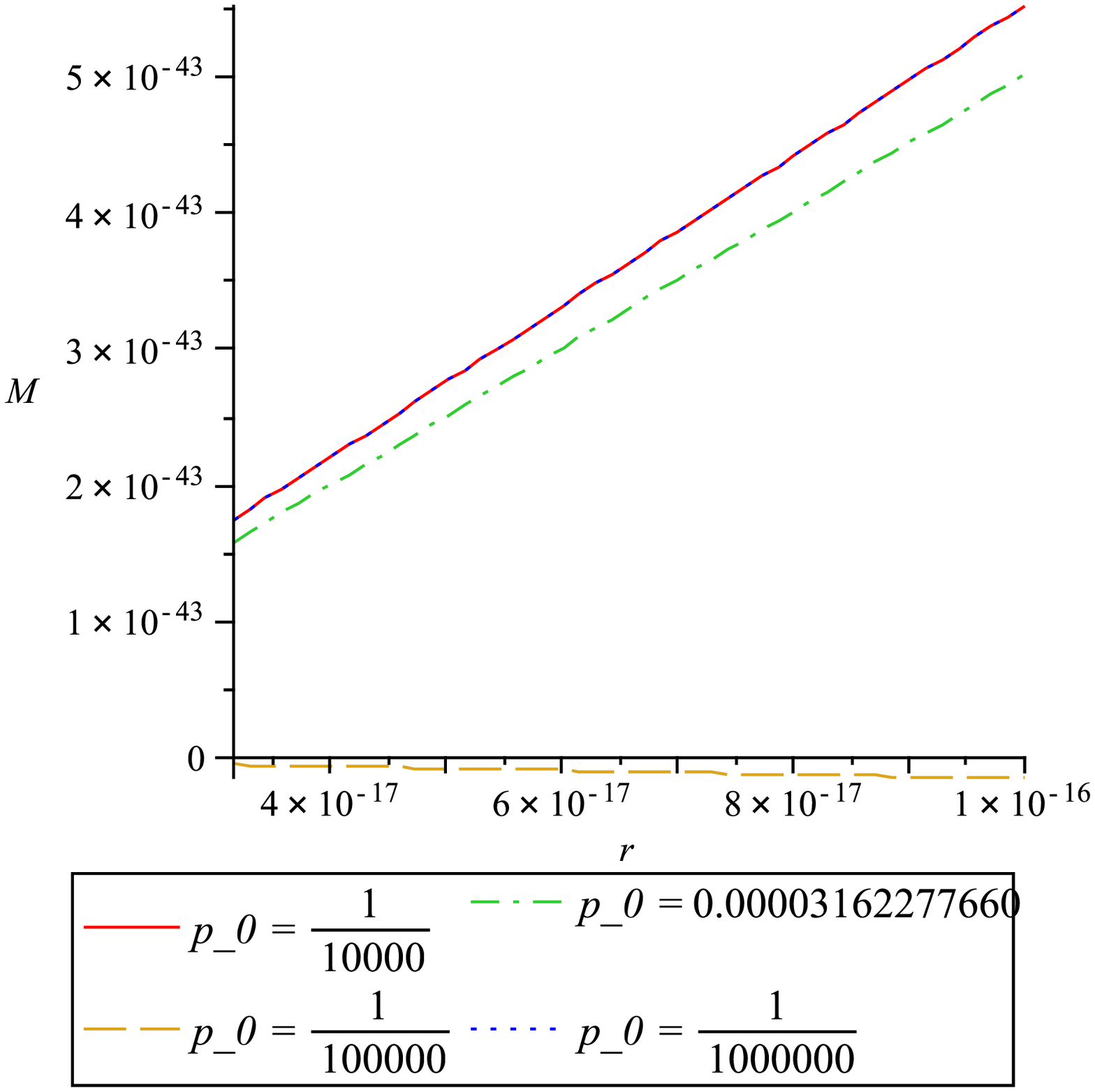}
\caption{\label{Fig.20}}
\end{figure}

\textbf{Case - II:}
\begin{equation}
\sigma e^{\frac{\lambda}{2}} = \sigma_{0} r^{s}
\end{equation}
($\sigma_{0}$ and $s$ are arbitrary constants)
Here we assume,
\begin{equation}
p_{r}(effective)=8\pi p-E^{2}=constant=p_{0},
\end{equation}
Here, the solutions are
\begin{equation}
p=\frac{1}{8\pi}\left[  p_{0}+\frac{16\pi^{2}\sigma_{0}^{2}}{(s+3)^{2}
}r^{2s+2}\right],
\end{equation}
\begin{equation}
\rho=\frac{-8\pi A}{\left[  p_{0}+\frac{16\pi^{2}\sigma_{0}^{2}}{(s+3)^{2}
}r^{2s+2}\right]  },
\end{equation}
\begin{equation}
\nu=\frac{2}{(s+1)}\ln\left[
\begin{array}
[c]{c}
-64A\pi^{2}+\\
+\left(  p_{0}+\frac{16\pi^{2}\sigma_{0}^{2}}{(s+3)^{2}}r^{2s+2}\right)  ^{2}
\end{array}
\right],
\end{equation}
\begin{equation}
e^{-\lambda}=1-\frac{2M(r)}{r},
\end{equation}
where,
\begin{widetext}
\begin{align}
M(r)=&\frac{8\pi^{2}
\sigma_{0}^{2}}{(2n+3)(n+2)^{2}}r^{2n+3}-\frac
{2A(n+2)^{2}}{n\xi^{2n-\eta}\sigma_{0}^{2}}\sum_{k=1}^{n}\sin\frac
{(2k-1)p\pi}{2n}tan^{-1}\left[  \frac{r+\xi\cos\frac{(2k-1)}{2n}\pi}{\xi
\sin\frac{(2k-1)}{2n}\pi}\right] \nonumber \\
&-\frac{1}{2n\xi^{2n-\eta}\sigma
_{0}^{2}}\sum_{k=1}^{n}\cos\frac{(2k-1)p\pi}{2n}\cdot
\ln\left[  r^{2}+2\xi r\cos\frac{(2k-1)}{2n}\pi+\xi^{2}\right].
\end{align}
\end{widetext}
[here, $0<\eta\leq2n$ and $n=s+1>0$( positive integer),$\xi=\frac
{(n+2)^{2}p_{0}}{16\pi^{2}\sigma_{0}^{2}}$, $\eta=3$] \newline
It can be observed that $8\pi p=E^{2}$ leads to a vanishing value of the
constant effective pressure term $p_{0}$ determining the corresponding
expressions for $p$, $\rho$, $\nu$, $e^{-\lambda}$ and $M(r)$ (Specialization
I, case II). Moreover, $p$ and $\rho$ have got $r^{2s+2}$ and $1/r^{2s+2}$
behaviour, respectively. For $8\pi p\neq E^{2}$ the fluid pressure is negative
for $p_{0}<-\frac{16\pi^{2}\sigma_{0}^{2}}{(s+3)^{2}}r^{2s+2}$. Further, for
$s=-3$ the model exhibits a singularity point for the fluid pressure $p$ and
the metric potential $\nu$, and the matter density $\rho$ tends toward zero.
The value $s=-1$ is a singularity for $\nu$. From $\rho=\frac{-8\pi A}{\left[
p_{0}+\frac{16\pi^{2}\sigma_{0}^{2}}{(s+3)}^{2}r^{2s+2}\right]  }$ it results
that there is no limiting case for WEC (only for $r\rightarrow\infty$), but
the matter density is positive for $p_{0}<-\frac{16\pi^{2}\sigma_{0}^{2}
}{(s+3)^{2}}r^{2s+2}$. Also, $p_{0}+\frac{16\pi^{2}\sigma_{0}^{2}}{(s+3)^{2}
}r^{2s+2}>\sqrt{64A\pi^{2}}$ (energy condition, limiting case is for equality
with zero) or $<\sqrt{64A\pi^{2}}$.

In Fig.21, Fig.22, Fig.23 and Fig.24 we plot $8\pi p$, $8\pi\rho$,
$e^{-\lambda}$ and $M(r)$ against $r$ for $s=0.5$.
\begin{figure}
\includegraphics[width=\columnwidth]{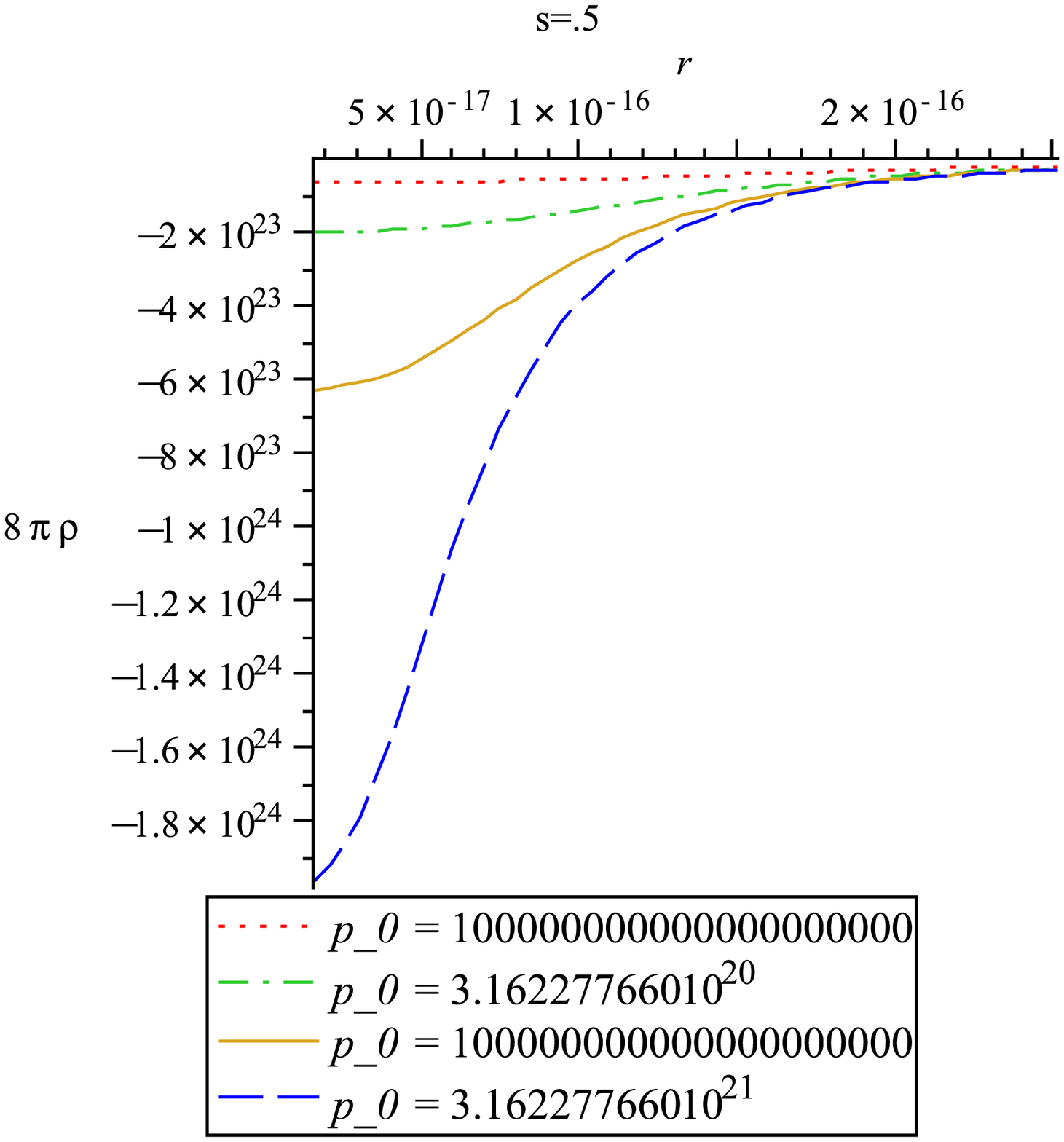}
\caption{\label{Fig.21}}
\end{figure}
\begin{figure}
\includegraphics[width=\columnwidth]{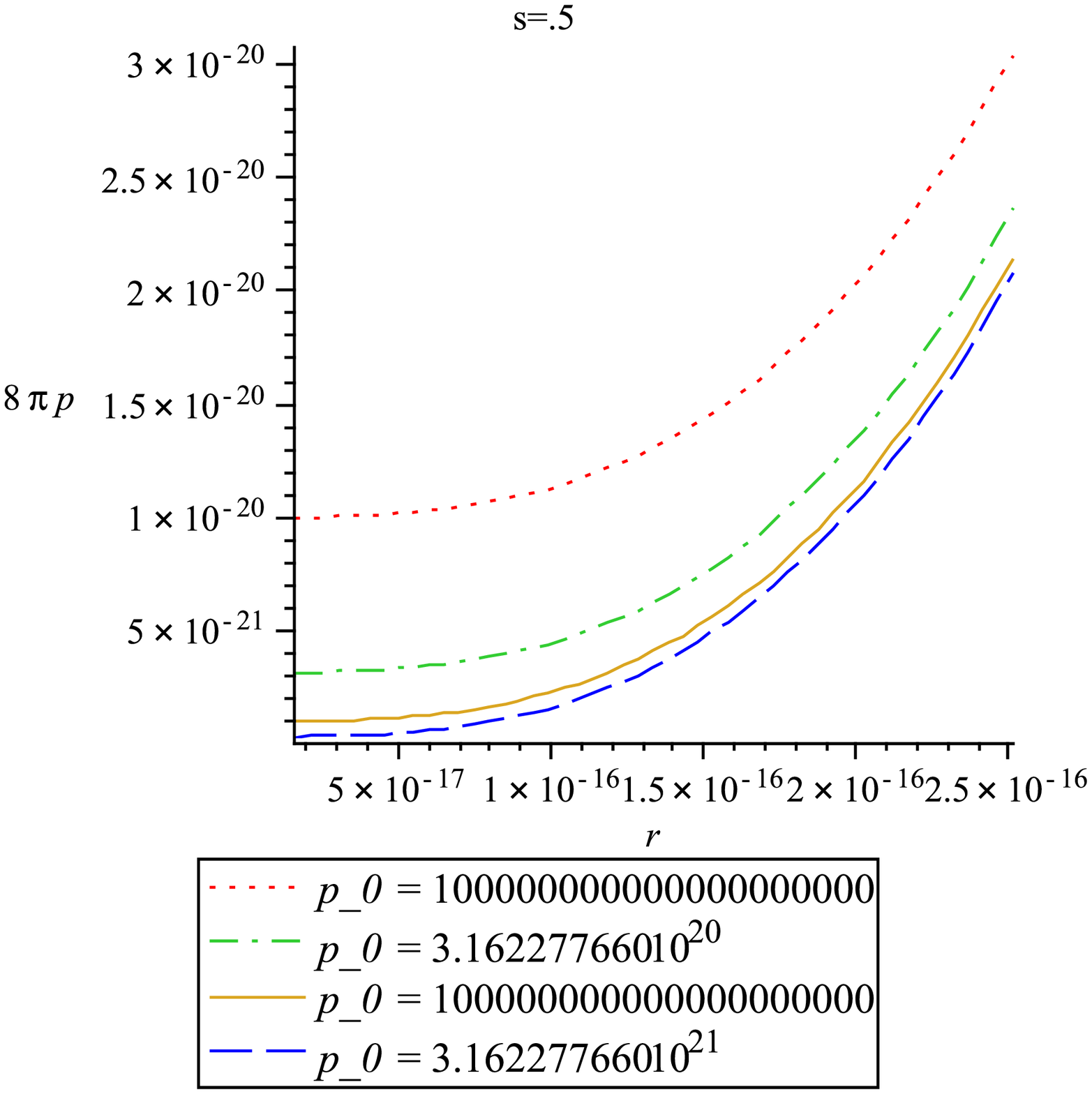}
\caption{\label{Fig.22}}
\end{figure}
\begin{figure}
\includegraphics[width=\columnwidth]{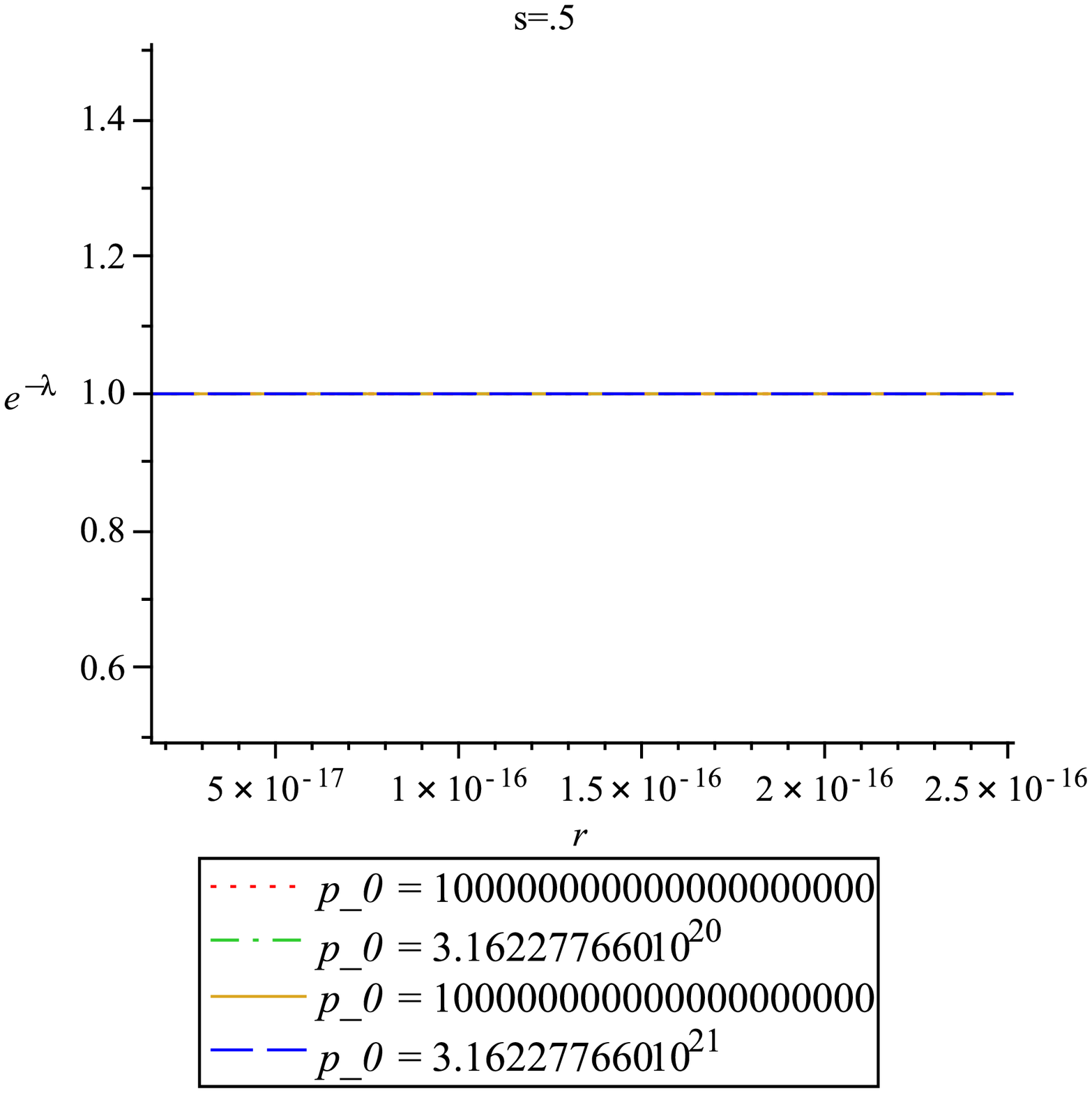}
\caption{\label{Fig.23}}
\end{figure}

Subcase I \textbf{$s=1$}

We obtain
\begin{equation}
p=\frac{1}{8\pi}\left[  p_{0}+\pi^{2}\sigma_{0}^{2}r^{4}\right]  ,
\end{equation}
\begin{equation}
\rho=\frac{-8\pi A}{\left[  p_{0}+\pi^{2}\sigma_{0}^{2}r^{4}\right]  },
\end{equation}
\begin{equation}
\nu=\ln\left[  -64A\pi^{2}+(p_{0}+\pi^{2}\sigma_{0}^{2}r^{4})^{2}\right]  ,
\end{equation}
\begin{equation}
e^{-\lambda}=1-\frac{2M(r)}{r},
\end{equation}
where
\begin{widetext}
\begin{align}
M(r) =\frac{\pi^{2}\sigma_{0}^{2}}{14}r^{7}-\frac{32A}{\sigma_{0}^{2}
}\left\{\frac{1}{4\sqrt{2}\alpha}\cdot \ln\left(\frac{r^{2}-\sqrt{2}\alpha r+\alpha^{2}}{r^{2}+\sqrt{2}\alpha
r+\alpha^{2}}\right)-\frac{1}{2\sqrt{2}\alpha}\cdot \left[\tan^{-1}(1-\frac{r\sqrt{2}}{\alpha})-\tan^{-1}(1+\frac
{r\sqrt{2}}{\alpha})\right]\right\},
\end{align}
\end{widetext}
with $\alpha^{4}=\frac{p_{0}}{\pi^{2}\sigma_{0}^{2}}$.
\begin{figure}
\includegraphics[width=\columnwidth]{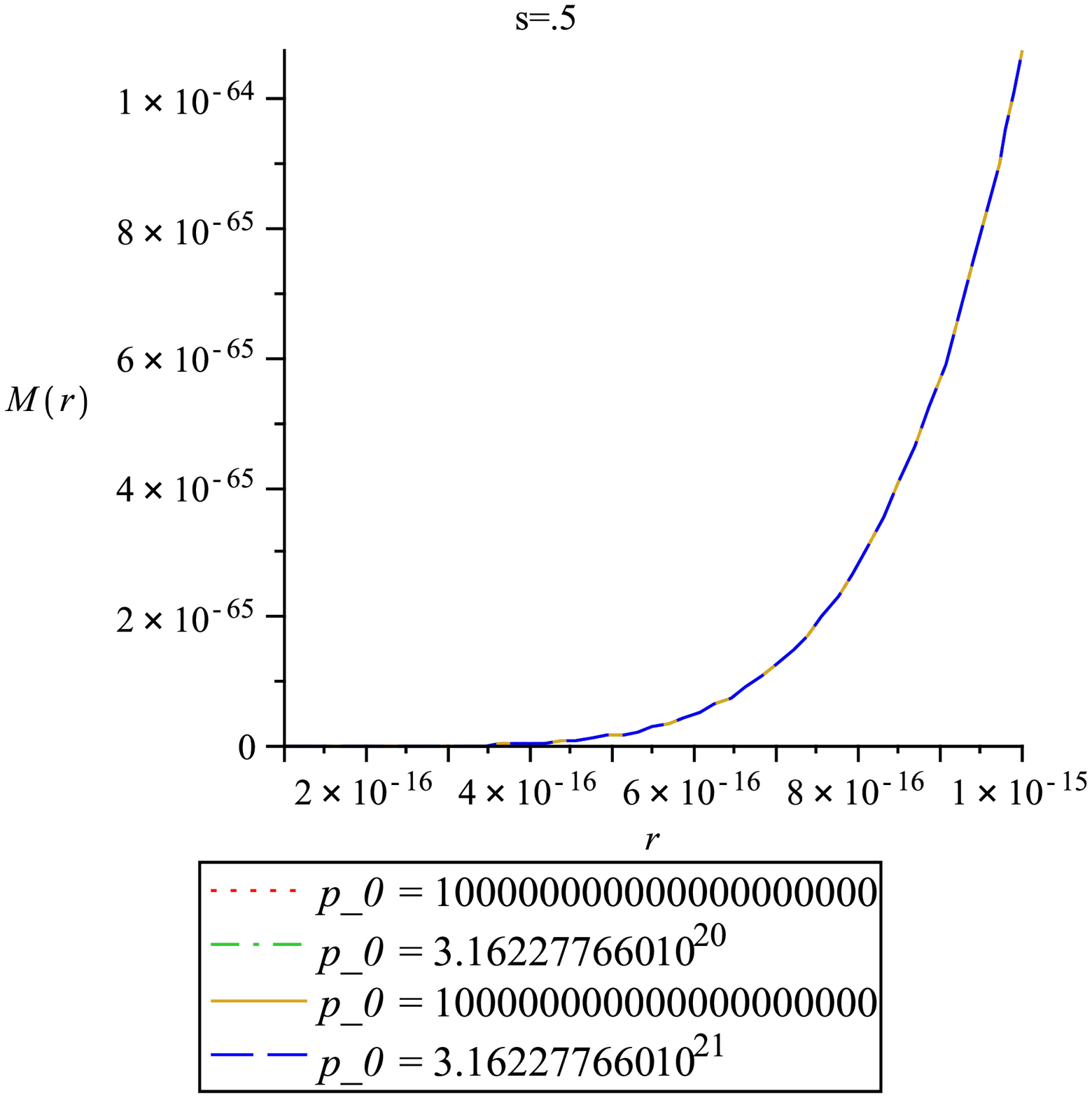}
\caption{\label{Fig.24}}
\end{figure}

For $8\pi p=E^{2}$ we get $p_{0}=0$, $p=\frac{\pi}{8}\sigma_{0}^{2}r^{4}$ and
$\rho=\frac{-8A}{\pi\sigma_{0}^{2}r^{4}}$. In the case $8\pi p\neq E^{2}$ we
notice that the pressure vanishes for $p_{0}=-\pi^{2}\sigma_{0}^{2}r^{4}$ and
is negative for $p_{0}<-\pi^{2}\sigma_{0}^{2}r^{4}$.

Subcase II \ \textbf{$s=\frac{1}{2}$}

We get
\begin{equation}
p=\frac{1}{8\pi}\left[  p_{0}+\frac{16\pi^{2}}{(3.5)^{2}}\sigma_{0}^{2}
r^{3}\right]  ,
\end{equation}
\begin{equation}
\rho=\frac{-8\pi A}{\left[  p_{0}+\frac{16\pi^{2}}{(3.5)^{2}}\sigma_{0}
^{2}r^{3}\right]  },
\end{equation}
\begin{equation}
\nu=\frac{4}{3}\ln\left[  -64A\pi^{2}+(p_{0}+\frac{16\pi^{2}}{(3.5)^{2}}
\sigma_{0}^{2}r^{3})^{2}\right]  ,
\end{equation}
\begin{equation}
e^{-\lambda}=1-\frac{2M(r)}{r},
\end{equation}
where
\begin{equation}
M(r)=\frac{4\pi^{2}\sigma_{0}^{2}r^{6}}{3(3.5)^{2}}-\frac{49}{6\sigma_{0}^{2}
}\ln[\frac{r^{3}+\beta^{3}}{\beta^{3}}],
\end{equation}
with $\beta^{3}=\frac{49p_{0}}{32\pi^{2}\sigma_{0}^{2}}$.

We notice that for $8\pi p=E^{2}$ we obtain $p=\frac{2\pi}{(3.5)^{2}}
\sigma_{0}^{2}r^{3}$ and $\rho=\frac{-8\pi A}{\frac{16\pi^{2}}{(3.5)^{2}
}\sigma_{0}^{2}r^{3}}$. For $8\pi p\neq E^{2}$ the fluid pressure is zero for
$p_{0}=-\frac{16\pi^{2}}{(3.5)^{2}}\sigma_{0}^{2}r^{3}$ and takes negative
values for $p_{0}<-\frac{16\pi^{2}}{(3.5)^{2}}\sigma_{0}^{2}r^{3}$. In this
case, also, at $r=0$, there is no singularity.

\section{Concluding Remarks}

We have developed some new toy electron gas models, and for this we have
considered a static spherically symmetric charged perfect fluid with the EOS
$p=-\frac{A}{\rho}$, where $A>0$, corresponding to the Chaplygin gas. For
performing the investigation our attention has been focused on three
specializations, each of them characterized by two particular cases. In
addition, for specialization III, case II the scenarios for the particular
values of the arbitrary constant $s=1$ and $s=\frac{1}{2}$ have been analyzed.
From our analysis we obtain the results:

Specialization I:

a) Case I: Assuming a vanishing value for the effective pressure
$p_{r}(effective)=0$, we obtain that $8\pi p=E^{2}$. The intensity of the
electric field, fluid pressure and charge density $\sigma_{0}$ ($\sigma$)
vanish due to a vanishing value of electric charge $q$. We get a new value for
the metric potential $\nu$ and $\sigma_{0}=0$ is a singularity point for the
matter density $\rho$, and for the active gravitational mass $M(r)$. We point
out that fluid pressure varies with respect of $p\thicksim r^{2}$ and the
matter density $\rho$ has got a $1/r^{2}$ behaviour. The limit $r\rightarrow
\infty$ connected to a non-zero value of $\sigma_{0}$ implies a zero value for
the matter density $\rho$, and infinity values for all the other physical
parameters. For this case we found isotropic coordinate as well as Kretschmann
scalar, further the Kretschmann scalar $R_{\mu\nu\alpha\beta}R^{\mu\nu
\alpha\beta}$ becomes infinity as $r\rightarrow0$.

b) Case II: In this case a vanishing value for $\sigma_{0}$ determines zero
values for the intensity of the electric field, electric charge and fluid
pressure, a modified value of the metric potential $\nu,$ and represents a
singularity point for the matter density $\rho$, $e^{-\lambda}$ and
gravitational mass $M(r)$. We observe that for the value $s=0$ we obtain the
Case I, for $s=-3$ the matter density $\rho$ vanishes and we get infinite
values for $E(r)$, $q(r)$, $p$, $\nu$, $M(r)$, and $s=\frac{1}{2}$,
$s=-\frac{5}{2}$ and $s=-1$ become singularity points for the gravitational
mass and metric potential $\nu$, respectively. In the limit $r\rightarrow0$,
$E(r)$, $q(r)$ and $p$ vanish, the fluid pressure becomes infinity with
respect $s>-1$ and the active gravitational mass become infinite for
$s>\frac{1}{2}$. For $\frac{2(s+3)^{2}A}{(2s-1)\sigma_{0}^{2}}r^{-2s+1}
=-\frac{8\pi^{2}\sigma_{0}^{2}}{(s+3)^{2}(2s+5)}r^{2s+5}$ with the
corresponding $s $ we obtain $m(a)=\frac{q^{2}}{2a}.$

Specialization II:

CaseI: $\ $From $8\pi\rho+E^{2}=constant=\rho_{0}$ it results that for
$8\pi\rho=-E^{2}$ the constant effective density term $\rho_{0}$ is zero. With
this condition, we obtain $\rho=-\frac{2\pi}{9}\sigma_{0}^{2}r^{2}$ and
$p=\frac{9A}{2\pi\sigma_{0}^{2}r^{2}}$, respectively. The gravitational mass
$M(r)$ vanishes, the metric potential $\nu$ has got a modified expression, and
the metric given by (2) contains a constant $g_{rr}$. In the case $8\pi
\rho\neq-E^{2}$ the matter density $\rho$ becomes positive and the fluid
pressure $p$ is negative with the condition $\rho_{0}>\frac{16\pi^{2}}
{9}\sigma_{0}^{2}r^{2}$ (compatible Chaplygin gas). The total gravitational
mass $m(r=a)$ is $m(a)=\frac{\rho_{0}}{6}a^{3}+\frac{8\pi^{2}\sigma_{0}
^{2}a^{5}}{9}$. The condition for obtaining a vanishing value for $\rho$ is
$\rho_{0}=E^{2}$. The negativity of the gravitational mass $\rho_{0}
<-\frac{16\pi^{2}}{3}\sigma_{0}^{2}r^{2}$ and energy density ($\rho<0$,
$\rho_{0}<\frac{16\pi^{2}}{9}\sigma_{0}^{2}r^{2}$) in the case of the electron
is connected with the Reissner-Nordstr\"{o}m repulsion. At the centre of the
spherical system we get $\rho=\frac{1}{8\pi}\rho_{0}$, which is constant and
$\nu=2\ln\left[  \frac{\rho_{0}}{\left[  64A\pi^{2}-\rho_{0}^{2}\right]  ^{2}
}\right]  $. The fluid pressure becomes negative according to $p=-\frac{8\pi
A}{\rho_{0}}$ (compatible Chaplygin gas), and this features is also obtained
for $\sigma_{0}=0$.

CaseII: The constant effective density term $\rho_{0}$ vanishes with respect
of $8\pi\rho=-E^{2}$. The matter-energy density $\rho$, pressure $p$, metric
potential $\nu$ and $e^{-\lambda}$ have finite values and for the active
gravitational mass we get a vanishing value. For a radius shrinking to the
center all the physical parameters from eqs. (52)-(54) are nonzero finite
quantities (the same as for $\sigma_{0}=0$) and the gravitational mass vanishes.

Specialization III:

Case I: In the limit $r\rightarrow0$ we obtain that $p$, $\rho$, $e^{\nu}$ and
$e^{\lambda}$ are all nonzero finite quantities i.e. there is no singularity
at the origin of the spherical system. For $8\pi p=E^{2}$ the constant
effective pressure $p_{0}$ vanishes. The case $8\pi p\neq E^{2}$ admits an
upper limit for the value of $p_{0}$ imposed by the condition of positivity of
the gravitational mass. The fluid pressure vanishes for $p_{0}=-\frac
{16\pi^{2}}{9}\sigma_{0}^{2}r^{2}$ or for the combined conditions $p_{0}=0$
($8\pi p=E^{2}$) and $\sigma_{0}=0$. Further, the fluid pressure takes
negative values for $p_{0}<-\frac{16\pi^{2}}{9}\sigma_{0}^{2}r^{2}$
(compatible Chaplygin model).

Case II: Note that from $8\pi p=E^{2}$ we get a vanishing value of $p_{0}$.
For $s=-3$ the model presents a singularity point for the fluid pressure $p$
and the metric potential $\nu$, with a vanishing value for the matter density
$\rho$. The value $s=-1$ represents a singularity for the metric potential
$\nu$. We obtain a negative value for the fluid pressure with respect
$p_{0}<-\frac{16\pi^{2}\sigma_{0}^{2}}{(s+3)^{2}}^{2}r^{2s+2}$ (compatible
Chaplygin model). Further, for $p_{0}=-\frac{16\pi^{2}\sigma_{0}^{2}
}{(s+3)^{2}}r^{2s+2}$ the fluid pressure vanishes. The value $s=0$ corresponds
to the Case I. Studying the particular cases $s=1$ and $s=\frac{1}{2}$ we
found that the fluid pressure vanishes for $p_{0}=-\pi^{2}\sigma_{0}^{2}r^{2}$
and $p_{0}=-\frac{16\pi^{2}\sigma_{0}^{2}}{(3.5)^{2}}r^{3}$, and take negative
values for $p_{0}<-\pi^{2}\sigma_{0}^{2}r^{2}$ and $p_{0}<-\frac{16\pi
^{2}\sigma_{0}^{2}}{(3.5)^{2}}r^{3}$, respectively.

In addition, a study of the WEC and dominant energy condition for all the
specializations has been performed.
\vspace*{-3mm}
\begin{acknowledgments}
\vspace*{-3mm}FR is thankful to Jadavpur University for financial support. We are also
grateful to Dr. V. Varela for several Illuminating discussions.
\end{acknowledgments}

\bibliography{ChaplyginREVTEX}

\end{document}